\documentclass[12pt]{article}

\usepackage{graphicx,booktabs,natbib,stmaryrd}
\usepackage{amsbsy, amstext, amssymb, amsthm, amsmath,bm}
\usepackage{algorithmic,algorithm}
\usepackage[margin=1in]{geometry}

\newtheorem{proposition}{{\bf Proposition}}
\newtheorem{lemma}{{\bf Lemma}}

\newtheorem{theorem}{{\bf Theorem}}

\newcommand{\vect}{\mathrm{vec}}

\newcommand{\real}[1]{\mathrm{I \! R} \mathit{^{#1}}}
\newcommand{\trans}{^{\mbox{\tiny {\sf T}}}}

\newcommand{\Bbf}{{\bm B}}

\newcommand{\Gbf}{{\bm G}}
\newcommand{\Hbf}{{\bm H}}
\newcommand{\Ibf}{{\bm I}}
\newcommand{\Jbf}{{\bm J}}

\newcommand{\Obf}{{\bm O}}
\newcommand{\Ubf}{{\bm U}}

\newcommand{\Xbf}{{\bm X}}
\newcommand{\Ybf}{{\bm Y}}
\newcommand{\Zbf}{{\bm Z}}

\newcommand{\bbf}{{\bm b}}

\newcommand{\xbf}{{\bm x}}

\newcommand{\greekbold}[1]{\mbox{\boldmath $#1$}}

\newcommand{\betabf}{\greekbold{\beta}}

\newcommand{\gammabf}{\greekbold{\gamma}}

\newcommand{\thetabf}{\greekbold{\theta}}

\newcommand{\Pibf}{\greekbold{\Pi}}

\bigskip
\title{\Large{\textbf{Tucker Tensor Regression\\}}
\Large{\textbf{and Neuroimaging Analysis}}}
\author{\medskip
\large{Xiaoshan Li, Hua Zhou and Lexin Li}\\
\large{North Carolina State University}
}
\date{}

\begin{document}
\maketitle

\begin{footnotetext}[1]
{\textit{Address for correspondence: Lexin Li, Department of Statistics, North Carolina State University, Box 8203, Raleigh, NC 27695-8203. Email: lexin\_li@ncsu.edu.}}
\end{footnotetext}

\baselineskip=21pt

\begin{abstract}
Large-scale neuroimaging studies have been collecting brain images of study individuals, which take the form of two-dimensional, three-dimensional, or higher dimensional arrays, also known as tensors. Addressing scientific questions arising from such data demands new regression models that take multidimensional arrays as covariates. Simply turning an image array into a long vector causes extremely high dimensionality that compromises classical regression methods, and, more seriously, destroys the inherent spatial structure of array data that possesses wealth of information. In this article, we propose a family of generalized linear tensor regression models based upon the Tucker decomposition of regression coefficient arrays. Effectively exploiting the low rank structure of tensor covariates brings the ultrahigh dimensionality to a manageable level that leads to efficient estimation. We demonstrate, both numerically that the new model could provide a  sound recovery of even high rank signals, and asymptotically that the model is consistently estimating the best Tucker structure approximation to the full array model in the sense of Kullback-Liebler distance. The new model is also compared to a recently proposed tensor regression model that relies upon an alternative CANDECOMP/PARAFAC (CP) decomposition.
\end{abstract}

\noindent{\bf Key Words:} CP decomposition; magnetic resonance image; tensor; Tucker decomposition.

\section{Introduction}
\label{sec:intro}

Advancing technologies are constantly producing large scale scientific data with complex structures. An important class arises from medical imaging, where the data takes the form of multidimensional array, also known as\ \emph{tensor}. Notable examples include electroencephalography (EEG, 2D matrix),  anatomical magnetic resonance images (MRI, 3D array), functional magnetic resonance images (fMRI, 4D array), among other image modalities. In medical imaging data analysis, a primary goal is to better understand associations between brains and clinical outcomes. Applications include using brain images to diagnose neurodegenerative disorders, to predict onset of neuropsychiatric diseases, and to identify disease relevant brain regions or activity patterns. This family of problems can collectively be formulated as a regression with clinical outcome as response, and image, or tensor, as predictor. However, the sheer size and complex structure of image covariate pose unusual challenges, which motivate us to develop a new class of regression models with image covariate. 

Most classical regression models take vector as covariate. Naively turning an image array into a vector is evidently unsatisfactory. For instance, a typical MRI image of size 128-by-128-by-128 implicitly requires $128^3 = 2,097,152$ regression parameters. Both computability and theoretical guarantee of the classical regression models are severely compromised by this ultra-high dimensionality. More seriously, vectorizing an array destroys the inherent spatial structure of the image array that usually possesses abundant information. A typical solution in the literature first employs the subject knowledge to extract a vector of features from images, and then feeds the feature vector into a classical regression model \citep{McKeown1998,Blankertz2001,Haxby2001,Kontos2003,Mitchell2004,Laconte2005,Ombao2006}. Alternatively one first applies unsupervised dimension reduction, often some variant of principal components analysis, to the image array, and then fits a regression model in the reduced dimensional vector space \citep{Caffo2010}. Both solutions are intuitive and popular, and have enjoyed varying degrees of success. At heart, both transform the problem to a classical vector covariate regression. However, there is no consensus on what choice best summarizes a brain image even for a single modality, whereas unsupervised dimension reduction like principal components could result in information loss in a regression setup. In contrast to constructing an image feature vector, the functional approach views image as a function and then employs functional regression models \citep{RamsaySilverman05FDABook}. \citet{ReissOgden10FunctionalGLM} notably applied this idea to regression with 2D image predictor. Extending their method to 3D and higher dimensional images, however, is far from trivial and requires substantial research, due to the large number of parameters and multi-collinearity among imaging measures.

In a recent work, \citet{ZhouLiZhu12CPTensor} proposed a class of generalized linear \emph{tensor} regression models. Specifically, for a response variable $Y$, a vector predictor $\Zbf \in \real{p_0}$ and a $D$-dimensional tensor predictor $\Xbf \in \real{p_1 \times \ldots \times p_D}$, the response is assumed to belong to an exponential family where the linear systematic part is of the form,
\begin{eqnarray} \label{eqn:glmlink}
    g(\mu) = \gammabf \trans \Zbf + \langle \Bbf,\Xbf \rangle.
\end{eqnarray}
Here $g(\cdot)$ is a strictly increasing link function, $\mu = E(Y | \Xbf, \Zbf)$, $\gammabf \in \real{p_0}$ is the regular regression coefficient vector, $\Bbf \in \real{p_1 \times \cdots \times p_D}$ is the coefficient array that captures the effects of tensor covariate $\Xbf$, and the inner product between two arrays is defined as $\langle \Bbf,\Xbf \rangle = \langle \vect \Bbf, \vect \Xbf \rangle = \sum_{i_1,\ldots,i_D} \beta_{i_1\ldots i_D} x_{i_1 \ldots i_D}$. This model, if with no further simplification, is prohibitive given its gigantic dimensionality: $p_0 + \prod_{d=1}^{D} p_d$. Motivated by a commonly used tensor decomposition,  \citet{ZhouLiZhu12CPTensor} introduced a low rank structure on the coefficient array $\Bbf$. That is, $\Bbf$ is assumed to follow a rank-$R$ CANDECOMP/PARAFAC (CP) decomposition \citep{KoldaBader09Tensor}, 
\begin{eqnarray} \label{eqn:cp}
    \Bbf = \sum_{r=1}^R \betabf_1^{(r)} \circ \cdots \circ \betabf_D^{(r)}, 
\end{eqnarray}
where $\betabf_d^{(r)} \in \real{p_d}$ are all column vectors, $d=1,\ldots,D,r=1,\ldots,R$, and $\circ$ denotes an outer product among vectors. Here the outer product $\bbf_1 \circ \bbf_2 \circ \cdots \circ \bbf_D$ of $D$ vectors $\bbf_d \in \real{p_d}$, $d=1,\ldots,D$, is defined as the $p_1 \times \cdots \times p_D$ array with entries $(\bbf_1 \circ \bbf_2 \circ \cdots \circ \bbf_D)_{i_1 \cdots i_D} = \prod_{d=1}^D b_{di_d}$. For convenience, this CP decomposition is often represented by a shorthand $\Bbf = \llbracket \Bbf_1,\ldots,\Bbf_D \rrbracket$, where $\Bbf_d = [\betabf_d^{(1)}, \ldots, \betabf_d^{(R)}] \in \real{p_d \times R}$, $d=1,\ldots,D$. Combining (\ref{eqn:glmlink}) and (\ref{eqn:cp}) yields generalized linear tensor regression models of \citet{ZhouLiZhu12CPTensor}, where the dimensionality decreases to the scale of $p_0 + R \times \sum_{d=1}^{D} p_d$. Under this setup, ultrahigh dimensionality of (\ref{eqn:glmlink}) is reduced to a manageable level, which in turn results in efficient estimation and prediction. For instance, for a regression with 128-by-128-by-128 MRI image and 5 usual covariates, the dimensionality is reduced from the order of $2,097,157 = 5 + 128^3$ to $389 = 5 + 128 \times 3$ for a rank-1 model, and to $1,157=5 + 3 \times 128 \times 3$ for a rank-3 model. \citet{ZhouLiZhu12CPTensor} showed that this low rank tensor model could provide a sound recovery of even high rank signals. 

In the tensor literature, there has been an important development parallel to CP decomposition, which is called Tucker decomposition, or higher-order singular value decomposition (HOSVD) \citep{KoldaBader09Tensor}. In this article, we propose a class of \emph{Tucker tensor regression models}. To differentiate, we call the models of \citet{ZhouLiZhu12CPTensor} \emph{CP tensor regression models}. Specifically, we continue to adopt the model (\ref{eqn:glmlink}), but assume that the coefficient array $\Bbf$ follows a Tucker decomposition, 
\begin{eqnarray} \label{eqn:tucker}
\Bbf = \sum_{r_1=1}^{R_1} \cdots \sum_{r_D=1}^{R_D} g_{r_1,\ldots,r_D} \betabf_{1}^{(r_1)} \circ \cdots \circ \betabf_{D}^{(r_D)}, 
\end{eqnarray}
where $\betabf_{d}^{(r_d)} \in \real{p_d}$ are all column vectors, $d=1,\ldots,D,r_d=1,\ldots,R_d$, and $g_{r_1,\ldots,r_D}$ are constants. It is often abbreviated as $\Bbf = \llbracket \Gbf; \Bbf_1,\ldots,\Bbf_D \rrbracket$, where $\Gbf \in \real{R_1 \times \cdots \times R_D}$ is a $D$-dimensional \emph{core tensor} with entries $(\Gbf)_{r_1\ldots r_D} = g_{r_1,\ldots,r_D}$, and $\Bbf_d \in \real{p_d \times R_d}$ are the factor matrices. $\Bbf_d$'s are usually orthogonal and can be thought of as the \emph{principal components} in each dimension (and thus the name, HOSVD). The number of parameters of a Tucker tensor model is in the order of $p_0 + \sum_{d=1}^{D} R_d \times p_d$. Comparing the two decompositions (\ref{eqn:cp}) and (\ref{eqn:tucker}), the key difference is that CP fixes the number of basis vectors $R$ along each dimension of $\Bbf$ so that all $\Bbf_d$'s have the \emph{same} number of columns (ranks). In contrast, Tucker allows the number $R_d$ to differ along different dimensions and $\Bbf_d$'s could have \emph{different} ranks.  

This difference between the two decompositions seems minor; however, in the context of tensor regression modeling and neuroimging analysis, it has profound implications, and such implications motivate this article. On one hand, the Tucker tensor regression model shares the advantages of the CP tensor regression model, in that it effectively exploits the special structure of the tensor data, it substantially reduces the dimensionality to enable efficient model estimation, and it provides a sound low rank approximation to a potentially high rank signal. On the other hand, Tucker tensor regression offers a much more \emph{flexible} modeling framework than CP regression, as it allows distinct order along each dimension. When the orders are all identical, it includes the CP model as a special case. This flexibility leads to several improvements that are particularly useful for neuroimaging analysis. First, a Tucker model could be more parsimonious than a CP model thanks to the flexibility of different orders. For instance, suppose a 3D signal $\Bbf \in \real{16 \times 16 \times 16}$ admits a Tucker decomposition \eqref{eqn:tucker} with $R_1=R_2=2$ and $R_3=5$. It can only be recovered by a CP decomposition with $R=5$, costing 230 parameters. In contrast, the Tucker model is more parsimonious with only 131 parameters. This reduction of free parameters is valuable for medical imaging studies, as the number of subjects is often limited. Second, the freedom in the choice of different orders is useful when the tensor data is skewed in dimensions, which is common in neuroimaging data. For instance, in EEG, the two dimensions consist of electrodes (channels) and time, and the number of sampling time points usually far exceeds the number of channels. Third, even when all tensor modes have comparable sizes, the Tucker formulation explicitly models the interactions between factor matrices $\Bbf_d$'s, and as such allows a finer grid search within a larger model space, which in turn may explain more trait variance. Finally, as we will show in Section 2.3, there exists a duality regarding the Tucker tensor model. Thanks to this duality, a Tucker tensor decomposition naturally lends itself to a principled way of imaging data downsizing, which, given the often limited sample size, again plays a practically very useful role in neuroimaging analysis. 

For these reasons, we feel it important to develop a complete methodology of Tucker tensor regression and its associated theory. The resulting Tucker tensor model carries a number of useful features. It performs dimension reduction through low rank tensor decomposition but in a supervised fashion, and as such avoids potential information loss in regression. It works for general array-valued image modalities and/or any combination of them, and for various types of responses, including continuous, binary, and count data. Besides, an efficient and highly scalable algorithm has been developed for the associated maximum likelihood estimation. This scalability is important considering the massive scale of imaging data. In addition, regularization has been studied in conjunction with the proposed model, yielding a collection of regularized Tucker tensor models, and particularly one that encourages sparsity of the core tensor to facilitate model selection among the defined Tucker model space. 

Recently there have been some increasing interests in matrix/tensor decomposition and their applications in brain imaging studies \citep{CrainiceanuCaffo11ImageDecomp,Allen2011MatDecomp,Hoff2011,AstonKirch2012}. Nevertheless, this article is distinct in that we concentrate on a regression framework with scalar response and tensor valued covariates. In contrast, \citet{CrainiceanuCaffo11ImageDecomp} and \citet{Allen2011MatDecomp} studied unsupervised decomposition, \citet{Hoff2011} considered model-based decomposition, whereas \citet{AstonKirch2012} focused on change point distribution estimation. The most closely related work to this article is \citet{ZhouLiZhu12CPTensor}; however, we feel our work is \emph{not} a simple extension of theirs. First of all, considering the complex nature of tensor, the development of the Tucker model estimation as well as its asymptotics is far from a trivial extension of the CP model of \citet{ZhouLiZhu12CPTensor}. Moreover, we offer a detailed comparison, both analytically (in Section \ref{sec:tucker-vs-cp}) and numerically (in Sections \ref{sec:num-compare} and \ref{sec:real-data}), of the CP and Tucker decompositions in the context of regression with imaging/tensor covariates. We believe this comparison is crucial for an adequate comprehension of tensor regression models and supervised tensor decomposition in general. 

The rest of the article is organized as follows. Section \ref{sec:model} begins with a brief review of some preliminaries on tensor, and then presents the Tucker tensor regression model. Section \ref{sec:estimation} develops an efficient algorithm for maximum likelihood estimation. Section \ref{sec:theory} derives inferential tools such as score, Fisher information, identifiability, consistency, and asymptotic normality. Section \ref{sec:regularization} investigates regularization method for the Tucker regression. Section \ref{sec:numerics} presents extensive numerical results. Section \ref{sec:discussion} concludes with some discussions and points to future extensions. All technical proofs are delegated to the Appendix.

\section{Model}
\label{sec:model}

\subsection{Preliminaries}

We start with a brief review of some matrix/array operations and results. Extensive references can be found in the survey paper \citep{KoldaBader09Tensor}. 

A \emph{tensor} is a multidimensional array. \emph{Fibers} of a tensor are the higher order analogue of matrix rows and columns. A fiber is
defined by fixing every index but one. A matrix column is a mode-1 fiber and a matrix row is a mode-2 fiber. Third-order tensors have column, row, and tube fibers, respectively. We next review some important operators that transform a tensor into a vector/matrix. The \emph{vec operator} stacks the entries of a $D$-dimensional tensor $\Bbf \in \real{p_1 \times \cdots \times p_D}$ into a column vector. Specifically, an entry $b_{i_1\ldots i_D}$ maps to the $j$-th entry of $\vect \, \Bbf$ where $j = 1 + \sum_{d=1}^D (i_d-1) \prod_{d'=1}^{d-1} p_{d'}$. For instance, when $D=2$, the matrix entry at cell $(i_1, i_2)$ maps to position $j= 1 + i_1 -1 + (i_2-1)p_1 = i_1 + (i_2-1)p_1$, which is consistent with the more familiar $\vect$ operator on a matrix. The \emph{mode-$d$ matricization}, $\Bbf_{(d)}$, maps a tensor $\Bbf$ into a $p_d \times \prod_{d' \ne d} p_{d'}$ matrix such that the $(i_1,\ldots,i_D)$ element of the array $\Bbf$ maps to the $(i_d,j)$ element of the matrix $\Bbf_{(d)}$, where $j = 1 + \sum_{d'\ne d} (i_{d'}-1) \prod_{d''<d',d'' \ne d} p_{d''}$. When $D=1$, we observe that $\vect \, \Bbf$ is the  same as vectorizing the mode-1 matricization $\Bbf_{(1)}$. The \emph{mode-($d,d'$) matricization} $\Bbf_{(dd')} \in \real{p_dp_{d'} \times \prod_{d'' \ne d,d'} p_{d''}}$ is defined in a similar fashion. We then define the \emph{mode-$d$ multiplication} of the tensor $\Bbf$ with a matrix $\Ubf \in \real{p_d \times q}$, denoted by $\Bbf \times_d \Ubf \in \real{p_1 \times \cdots \times q \times \cdots \times p_D}$, as the multiplication of the mode-$d$ fibers of $\Bbf$ by $\Ubf$. In other words, the mode-$d$ matricization of $\Bbf \times_d \Ubf$ is $\Ubf \Bbf_{(d)}$. 

We also review two properties of a tensor $\Bbf$ that admits a Tucker decomposition (\ref{eqn:tucker}). The mode-$d$ matricization of $\Bbf$ can be expresses as
\begin{eqnarray*} \label{eqn:tucker-matricization}
    \Bbf_{(d)} = \Bbf_d \Gbf_{(d)} (\Bbf_D \otimes \cdots \otimes \Bbf_{d+1} \otimes \Bbf_{d-1} \otimes \cdots \otimes \Bbf_1) \trans,
\end{eqnarray*}
where $\otimes$ denotes the Kronecker product of matrices. If applying the $\vect$ operator to $\Bbf$, then 
\begin{eqnarray*} \label{eqn:tucker-vec}
    \vect \Bbf = \vect \Bbf_{(1)} = \vect ( \Bbf_1 \Gbf_{(1)} (\Bbf_D \otimes \cdots \otimes \Bbf_2) \trans) = (\Bbf_D \otimes \cdots \otimes \Bbf_1) \vect \Gbf. 
\end{eqnarray*}
These two properties are useful for our subsequent Tucker regression development.

\subsection{Tucker Regression Model}

We elaborate on the Tucker tensor regression model introduced in Section 1. We assume that $Y$ belongs to an exponential family with probability mass function or density \citep{McCullaghNelder83GLMBook}, 
\begin{eqnarray*} \label{eqn:GLM-density}
p(y_i|\theta_i,\phi) = \exp\left\{ \frac{y_i\theta_i - b(\theta_i)}{a(\phi)} + c(y_i,\phi) \right\} 
\end{eqnarray*}
with the first two moments $E(Y_i) = \mu_i = b'(\theta_i)$ and $\mathrm{Var}(Y_i) = \sigma_i^2 = b''(\theta_i) a_i(\phi)$. $\theta$  and $\phi>0$ are, respectively, called the natural and dispersion parameters. We assume the systematic part of GLM is of the form
\begin{eqnarray} \label{eqn:r-tensorreg}
    g(\mu) = \eta = \gammabf \trans \Zbf +  \langle \sum_{r_1=1}^{R_1} \cdots \sum_{r_D=1}^{R_D} g_{r_1,\ldots,r_D} \betabf_{1}^{(r_1)} \circ \cdots \circ \betabf_{D}^{(r_D)},\Xbf  \rangle. 
\end{eqnarray}
That is, we impose a Tucker structure on the array coefficient $\Bbf$. We make a few remarks. First, in this article, we consider the problem of estimating the core tensor $\Gbf$ and factor matrices $\Bbf_d$ simultaneously given the response $Y$ and covariates $\Xbf$ and $\Zbf$. This can be viewed as a \emph{supervised} version of the classical unsupervised Tucker decomposition. It is also a supervised version of principal components analysis for higher-order multidimensional array. Unlike a two-stage solution that first performs principal components analysis and then fits a regression model, the basis (principal components) $\Bbf_d$ in our models are estimated under the guidance (supervision) of the response variable. Second, the CP model of \citet{ZhouLiZhu12CPTensor} corresponds to a special case of the Tucker model \eqref{eqn:r-tensorreg} with $g_{r_1,\ldots,r_D}= 1_{\{r_1 = \cdots =r_D\}}$ and $R_1=\ldots=R_D=R$. In other words, the CP model is a specific Tucker model with a super-diagonal core tensor $\Gbf$. The CP model has a rank at most $R$ while the general Tucker model can have a rank as high as $R^D$. We will further compare the two model sizes in Section \ref{sec:tucker-vs-cp}.

\subsection{Duality and Tensor Basis Pursuit}
\label{sec:duality}

Next we investigate a duality regarding the inner product between a general tensor and a tensor that admits a Tucker decomposition. 

\begin{lemma}[Duality]
\label{lemma:duality}
Suppose a tensor $\Bbf \in \real{p_1 \times \cdots \times p_D}$ admits Tucker decomposition $\Bbf = \llbracket \Gbf; \Bbf_1, \ldots, \Bbf_D \rrbracket$. Then, for any tensor $\Xbf \in \real{p_1 \times \cdots \times p_D}$, $\langle \Bbf, \Xbf \rangle = \langle \Gbf, \tilde \Xbf \rangle$, where $\tilde \Xbf$ admits a Tucker decomposition $\tilde \Xbf = \llbracket \Xbf; \Bbf_1 \trans, \ldots, \Bbf_D \trans \rrbracket$.
\end{lemma}
\noindent
This duality gives some important insights to the Tucker tensor regression model. First, if we consider $\Bbf_d \in \real{p_d \times R_d}$ as fixed and known basis matrices, then Lemma \ref{lemma:duality} says fitting the Tucker tensor regression model \eqref{eqn:r-tensorreg} is equivalent to fitting a tensor regression model in $\Gbf$ with the \emph{transformed} data $\tilde \Xbf = \llbracket \Xbf; \Bbf_1 \trans,\ldots,\Bbf_D \trans \rrbracket \in \real{R_1 \times \cdots \times R_D}$. When $R_d \ll p_d$, the transformed data $\tilde \Xbf$ effectively \emph{downsize} the original data. We will further illustrate this downsizing feature in the real data analysis in Section \ref{sec:real-data}. Second, in applications where the numbers of basis vectors $R_d$ are unknown, we can utilize possibly over-complete basis matrices $\Bbf_d$ such that $R_d \ge p_d$, and then estimate $\Gbf$ with sparsity regularizations. This leads to a tensor version of the classical basis pursuit problem \citep{ChenDonohoSaunders01BasisPursuit}. Take fMRI data as an example. We can adopt the wavelet basis for the three image dimensions and the Fourier basis for the time dimension. Regularization on $\Gbf$ can be achieved by either imposing a low rank decomposition (CP or Tucker) on $\Gbf$ (hard thresholding) or penalized regression (soft thresholding). We will investigate Tucker regression regularization in details in Section \ref{sec:regularization}.

\subsection{Model Size: Tucker vs CP}
\label{sec:tucker-vs-cp}

In this section we investigate the size of the Tucker tensor model. Comparison with the size of the CP tensor model helps gain better understanding of both models. In addition, it provides a base for data adaptive selection of appropriate orders in a Tucker model.

First we quickly review the number of free parameters $p_{\text{C}}$ for a CP model $\Bbf = \llbracket \Bbf_1, \ldots, \Bbf_d \rrbracket$, with $\Bbf_d \in \real{p_d \times R}$. For $D=2$, $p_{\text{C}} = R(p_1+p_2)-R^2$, and for $D > 2$, $p_{\text{C}} = R(\sum_{d=1}^D p_d - D + 1)$. For $D=2$, the term $-R^2$ adjusts for the nonsingular transformation indeterminacy for model identifiability; for $D>2$, the term $R(-D+1)$ adjusts for the scaling indeterminacy in the CP decomposition. See \citet{ZhouLiZhu12CPTensor} for more details. Following similar arguments, we obtain that the number of free parameters $p_{\text{T}}$ in a Tucker model $\Bbf = \llbracket \Gbf; \Bbf_1, \ldots, \Bbf_d \rrbracket$, with $\Gbf \in \real{R_1 \times \cdots \times R_d}$ and $\Bbf_d \in \real{p_d \times R_d}$, is 
\begin{eqnarray*}
p_{\text{T}} = \sum_{d=1}^D p_dR_d + \prod_{d=1}^D R_d - \sum_{d=1}^D R_d^2,
\end{eqnarray*}
for any $D$. Here the term -$\sum_{d=1}^D R_d^2$ adjusts for the non-singular transformation indeterminancy in the Tucker decomposition. We summarize these results in Table \ref{tab:model-size}. 

Next we compare the two model sizes (degrees of freedom) under an additional assumption that $R_1 = \cdots = R_d = R$. The difference becomes:
\begin{align*}
    p_{\text{T}} - p_{\text{C}} = \begin{cases}
    0    & \textrm{ when } D=2,   \\
    R(R-1)(R-2) & \textrm{ when } D=3,    \\
    R(R^3-4R+3) & \textrm{ when } D=4,    \\
    R(R^{D-1}-DR+D-1) & \textrm{ when } D>4.
    \end{cases}
\end{align*}
Based on this formula, when $D=2$, the Tucker model is essentially the same as the CP model. When $D=3$, Tucker has the same number of parameters as CP for $R=1$ or $R=2$, but costs $R(R-1)(R-2)$ more parameters for $R>2$. When $D>3$, Tucker and CP are the same for $R=1$, but Tucker costs substantially more parameters than CP for $R>2$. For instance, when $D=4$ and $R=3$, Tucker model takes 54 more parameters than the CP model. However, one should bear in mind that the above discussion assumes $R_1=\cdots=R_d=R$. In reality, Tucker could require \emph{less} free parameters than CP, as shown in the illustrative example given in Section 1, since Tucker is more flexible and allows different order $R_d$ along each dimension. 

\begin{table}[t]
\caption{Number of free parameters in Tucker and CP models.}
\label{tab:model-size}
\vspace{-0.1in}
\begin{center}
\begin{tabular}{|c|cc|} \hline
& CP & Tucker   \\ \hline
$D=2$ & $R(p_1+p_2)-R^2$ & $p_1R_1 + p_2R_2+R_1R_2-R_1^2 -R_2^2$ \\
$D>2$ & $R(\sum_d p_d -D+1)$ & $\sum_d p_dR_d + \prod_d R_d - \sum_d R_d^2$ \\ \hline
\end{tabular}
\end{center}
\end{table}

\begin{figure}[t]
\begin{center}
$$
\begin{array}{cc}
\includegraphics[width=2.3in]{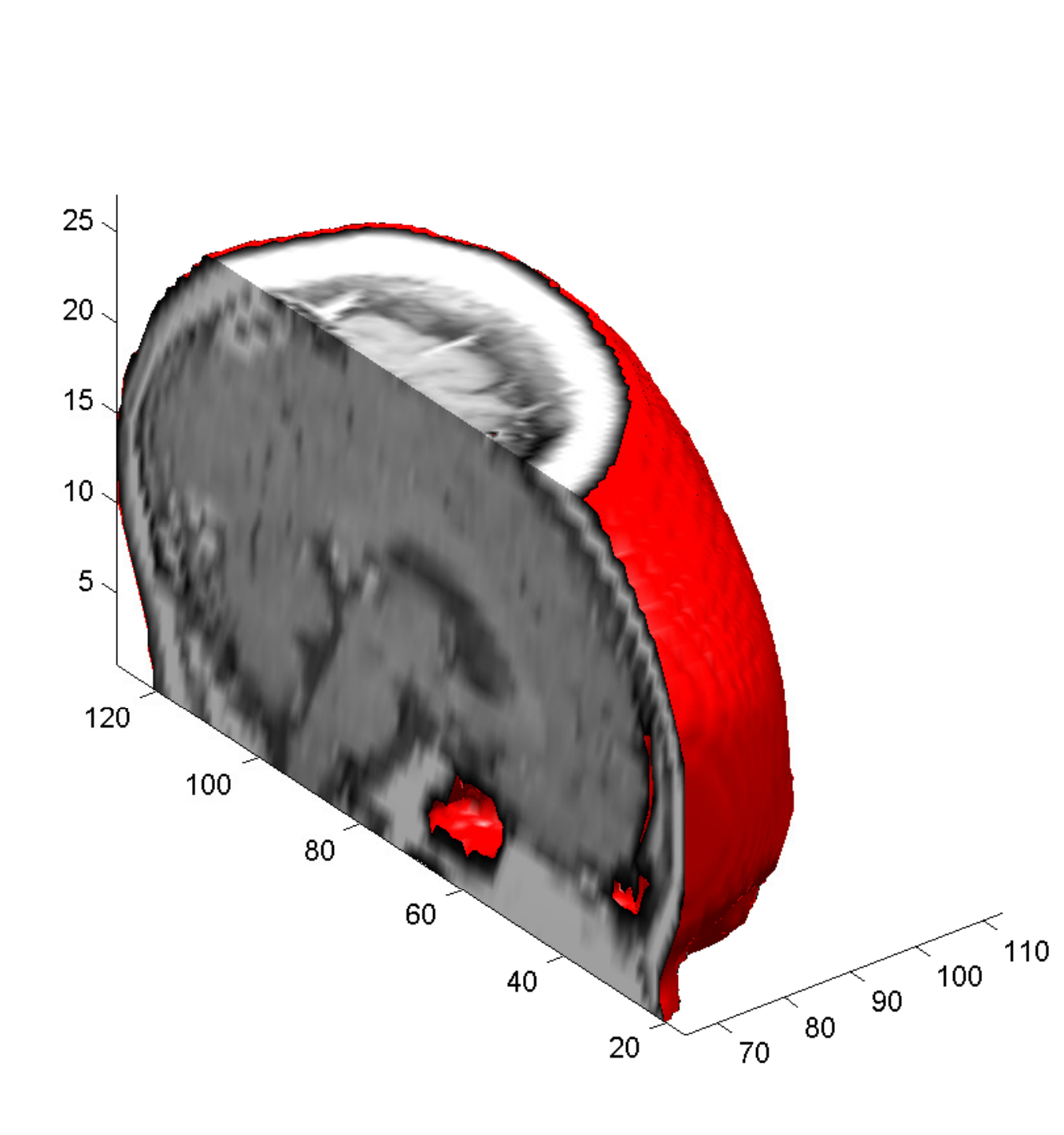}  & \includegraphics[width=2.3in]{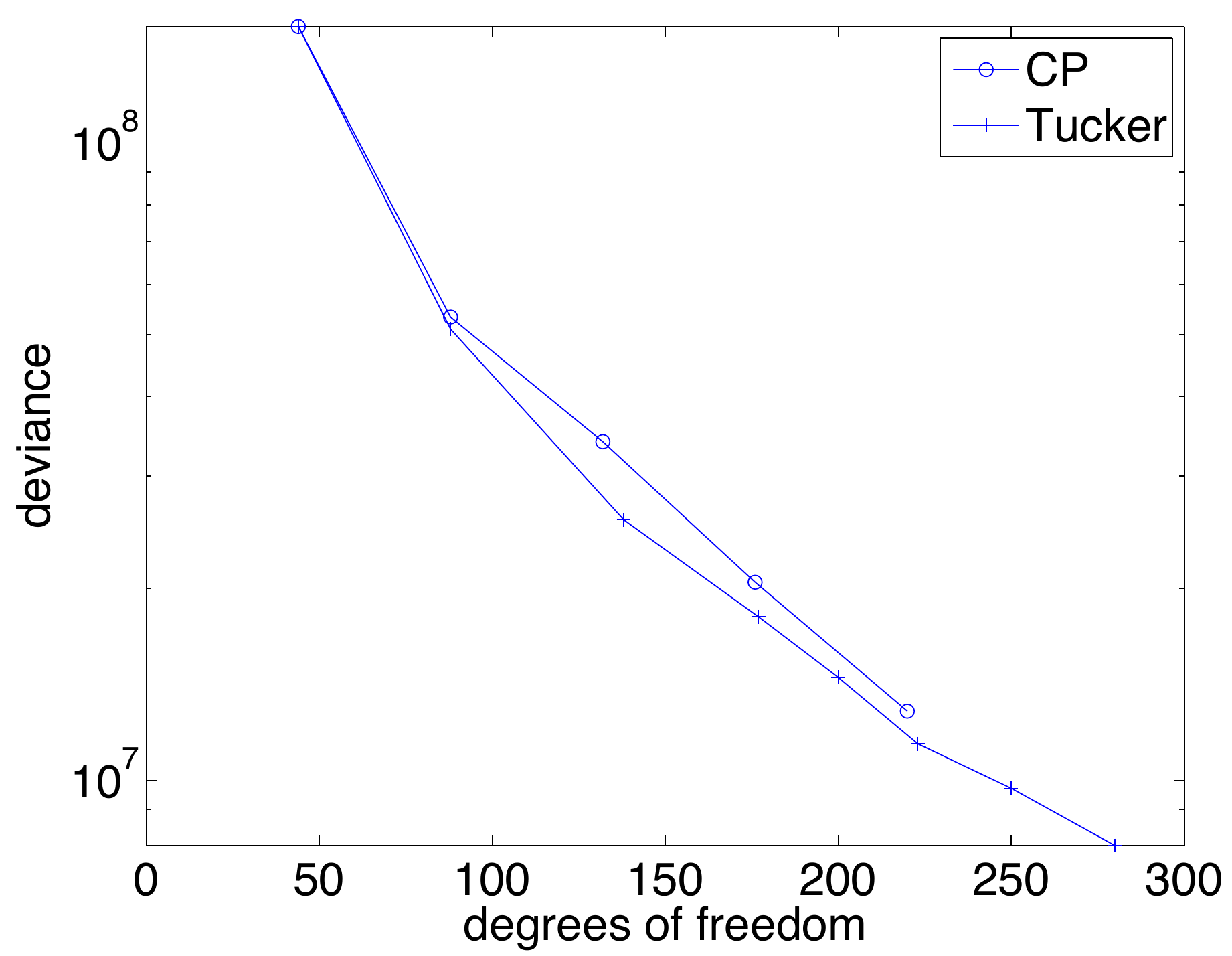}
\end{array}
$$
\end{center}
\caption{Left: half of the true signal array $\Bbf$. Right: Deviances of CP regression estimates at $R=1,\ldots,5$, and Tucker regression estimates at orders $(R_1,R_2,R_3)=(1,1,1)$, $(2, 2, 2)$, $(3, 3, 3)$, $(4, 4, 3)$, $(4, 4, 4)$, $(5, 4, 4)$, $(5, 5, 4)$, and $(5, 5, 5)$. The sample size is $n=1000$.}
\label{fig:Tucker-vs-Kruskal-skull}
\end{figure}

Figure \ref{fig:Tucker-vs-Kruskal-skull} shows an example with $D=3$ dimensional array covariates. Half of the true signal (brain activity map) $\Bbf$ is displayed in the left panel, which is by no means a low rank signal. Suppose 3D images $\Xbf_i$ are taken on $n=1,000$ subjects. We simulate image traits $\Xbf_i$ from independent standard normals and quantitative traits $Y_i$ from independent normals with mean $\langle \Xbf_i, \Bbf \rangle$ and unit variance. Given the limited sample size, the hope is to infer a reasonable low rank approximation to the activity map from the 3D image covariates. The right panel displays the model deviance versus the degrees of freedom of a series of CP and Tucker model estimates. The CP model is estimated at ranks $R=1,\ldots,5$. The Tucker model is fitted at orders $(R_1,R_2,R_3)=(1,1,1)$, $(2, 2, 2)$, $(3, 3, 3)$, $(4, 4, 3)$, $(4, 4, 4)$, $(5, 4, 4)$, $(5, 5, 4)$, and $(5, 5, 5)$. We see from the plot that, under the same number of free parameters, the Tucker model could generally achieve a better model fit with a smaller deviance. (Note that the deviance is in the log scale, so a small discrepancy between the two lines translates to a large value of difference in deviance.) 

The explicit model size formula of the Tucker model is also useful for choosing appropriate orders $R_d$'s along each direction given data. This can be treated as a model selection problem, and we can employ a typical model selection criterion, e.g., Bayesian information criterion (BIC). It is of the form: $-2 \log \ell + \log(n) p_e$, where $\ell$ is the log-likelihood, and $p_e = p_{\text{T}}$ is the effective number of parameters of the Tucker model as given in Table ~\ref{tab:model-size}. We will illustrate this BIC criterion in the numerical Section \ref{sec:shapes}, and will discuss some heuristic guidelines of selecting orders in Section \ref{sec:real-data}.

\section{Estimation}
\label{sec:estimation}

We pursue the maximum likelihood estimation (MLE) for the Tucker tensor regression model and develop a scalable estimation algorithm in this section. The key observation is that, although the systematic part \eqref{eqn:r-tensorreg} is not linear in $\Gbf$ and $\Bbf_d$ \emph{jointly}, it is linear in them \emph{separately}. This naturally suggests a block relaxation algorithm, which updates each factor matrix $\Bbf_d$ and the core tensor $\Gbf$ \emph{alternately}. 

The algorithm consists of two core steps. First, when updating $\Bbf_d \in \real{p_d \times R_d}$ with the rest $\Bbf_{d'}$'s and $\Gbf$ fixed , we rewrite the array inner product in (\ref{eqn:r-tensorreg}) as
\begin{eqnarray*}
\langle \Bbf,\Xbf \rangle 
    &=& \langle \Bbf_{(d)},\Xbf_{(d)} \rangle \\
    &=& \langle \Bbf_d \Gbf_{(d)} (\Bbf_D \otimes \cdots \otimes \Bbf_{d+1} \otimes \Bbf_{d-1} \otimes \cdots \otimes \Bbf_1) \trans, \Xbf_{(d)} \rangle  \\
    &=& \langle \Bbf_d, \Xbf_{(d)} (\Bbf_D \otimes \cdots \otimes \Bbf_{d+1} \otimes \Bbf_{d-1} \otimes \cdots \otimes \Bbf_1) \Gbf_{(d)} \trans \rangle.
\end{eqnarray*}
Then the problem turns into a GLM regression with $\Bbf_d$ as the ``parameter" and the term $\Xbf_{(d)} (\Bbf_D \otimes \cdots \otimes \Bbf_{d+1} \otimes \Bbf_{d-1} \otimes \cdots \otimes \Bbf_1) \Gbf_{(d)} \trans$ as the ``predictor". It is a low dimensional GLM with only $p_dR_d$ parameters and thus is easy to solve. Second, when updating $\Gbf \in \real{R_1 \times \cdots \times R_D}$ with all $\Bbf_{d}$'s fixed,
\begin{eqnarray*}
    \langle \Bbf,\Xbf \rangle & = & \langle \vect \Bbf, \vect \Xbf \rangle \\
    &=& \langle (\Bbf_D \otimes \cdots \otimes \Bbf_1) \vect \Gbf, \vect \Xbf \rangle \\
    &=& \langle \vect \Gbf, (\Bbf_D \otimes \cdots \otimes \Bbf_1) \trans \vect \Xbf \rangle.
\end{eqnarray*}
This implies a GLM regression with $\vect \Gbf$ as the ``parameter" and the term $(\Bbf_D \otimes \cdots \otimes \Bbf_1) \trans \vect \Xbf$ as the "predictor". Again this is a low dimensional regression problem with $\prod_d R_d$ parameters. For completeness, we summarize the above alternating estimation procedure in Algorithm \ref{algo:br-algo}. The orthogonality between the columns of factor matrices $\Bbf_d$ is not enforced as in unsupervised HOSVD, because our primary goal is approximating tensor signal instead of finding the principal components along each mode.

\begin{algorithm}
\begin{algorithmic}
\STATE Initialize: $\gammabf^{(0)} = \mbox{argmax}_{\gammabf} \, \ell(\gammabf,{\bf 0}, \ldots, {\bf 0})$, $\Bbf_d^{(0)} \in $ $\real{p_d \times R_d}$ a random matrix for $d=1,\ldots,D$, and $\Gbf^{(0)} \in \real{R_1 \times \cdots \times R_D}$ a random matrix.
\REPEAT
\FOR{$d=1, \ldots, D$}
\STATE $\Bbf_d^{(t+1)} = \mbox{argmax}_{\Bbf_d} \, \ell(\gammabf^{(t)}, \Bbf_1^{(t+1)}, \ldots, \Bbf_{d-1}^{(t+1)}, \Bbf_d, \Bbf_{d+1}^{(t)}, \ldots, \Bbf_D^{(t)},\Gbf^{(t)})$
\ENDFOR
\STATE $\Gbf^{(t+1)} = \mbox{argmax}_{\Gbf} \, \ell(\gammabf^{(t)}, \Bbf_1^{(t+1)}, \ldots, \Bbf_D^{(t+1)},\Gbf)$
\STATE $\gammabf^{(t+1)} = \mbox{argmax}_{\gammabf} \, \ell(\gammabf, \Bbf_1^{(t+1)}, \ldots, \Bbf_D^{(t+1)},\Gbf^{(t+1)})$
\UNTIL{$\ell(\thetabf^{(t+1)})-\ell(\thetabf^{(t)}) < \epsilon$}
\end{algorithmic}
\caption{Block relaxation algorithm for fitting the Tucker tensor regression.}
\label{algo:br-algo}
\end{algorithm}

Next we study the convergence properties of the proposed algorithm. As the block relaxation algorithm monotonically increases the objective value, the stopping criterion is well-defined and the convergence properties of iterates follow from the standard theory for monotone algorithms \citep{deLeeuw94BR,Lange10NumAnalBook}. The proof of next result is given in the Appendix.
\begin{proposition}
\label{prop:glob-conv}
Assume (i) the log-likelihood function $\ell$ is continuous, coercive, i.e., the set $\{\thetabf: \ell(\thetabf) \ge \ell(\thetabf^{(0)})\}$ is compact, and bounded above, (ii) the objective function in each block update of Algorithm \ref{algo:br-algo} is strictly concave, and (iii) the set of stationary points (modulo nonsingular transformation indeterminacy) of $\ell(\gammabf,\Gbf,\Bbf_1,\ldots,\Bbf_D)$ are isolated. We have the following results.
\begin{enumerate}
\item (Global Convergence) The sequence $\thetabf^{(t)} = (\gammabf^{(t)}, \Gbf^{(t)}, \Bbf_1^{(t)}, \ldots, \Bbf_D^{(t)})$ generated by Algorithm \ref{algo:br-algo} converges to a stationary point of $\ell(\gammabf,\Gbf,\Bbf_1,\ldots,\Bbf_D)$.
\item (Local Convergence) Let $\thetabf^{(\infty)} = (\gammabf^{(\infty)},\Gbf^{(\infty)},\Bbf_1^{(\infty)},\ldots,\Bbf_D^{(\infty)})$ be a strict local maximum of $\ell$. The iterates generated by Algorithm \ref{algo:br-algo} are locally attracted to $\thetabf^{(\infty)}$ for $\thetabf^{(0)}$ sufficiently close to $\thetabf^{(\infty)}$.
\end{enumerate}
\end{proposition}

\section{Statistical Theory}
\label{sec:theory}

In this section we study the usual large $n$ asymptotics of the proposed Tucker tensor regression. Regularization is treated in the next section for the small or moderate $n$ cases. For simplicity, we drop the classical covariate $\Zbf$ in this section, but all the results can be straightforwardly extended to include $\Zbf$. We also remark that, although the usually limited sample size of neuroimging studies makes the large $n$ asymptotics seem irrelevant, we still believe such an asymptotic investigation important, for several reasons. First, when the sample size $n$ is considerably larger than the effective number of parameters $p_{\text{T}}$, the asymptotic study tells us that the model is consistently estimating the best Tucker structure approximation to the full array model in the sense of Kullback-Liebler distance. Second, the explicit formula for score and information are not only useful for asymptotic theory but also for computation, while the identifiability issue has to be properly dealt with for the given model. Finally, the regular asymptotics can be of practical relevance, for instance, can be useful in a likelihood ratio type test in a replication study.

\subsection{Score and Information}

We first derive the score and information for the tensor regression model, which are essential for statistical estimation and inference. The following standard calculus notations are used. For a scalar function $f$, $\nabla f$ is the (column) gradient vector, $df = [\nabla f] \trans$ is the differential, and $d^2f$ is the Hessian matrix. For a multivariate function $g: \real{p} \mapsto \real{q}$, $Dg \in \real{p \times q}$ denotes the Jacobian matrix holding partial derivatives $\frac{\partial g_j}{\partial x_i}$. We start from the Jacobian and Hessian of the systematic part $\eta \equiv g(\mu)$ in (\ref{eqn:r-tensorreg}).
\begin{lemma}
\label{lemma:eta-derivatives}
\begin{enumerate}
\item The gradient $\nabla \eta(\Bbf_1,\ldots,\Bbf_D) \in \real{\prod_d R_d + \sum_{d=1}^D p_d R_d}$ is
\begin{eqnarray*}
    \nabla \eta(\Gbf, \Bbf_1,\ldots,\Bbf_D) = [\Bbf_D \otimes \cdots \otimes \Bbf_1 \,\, \Jbf_1 \,\, \Jbf_2 \,\, \cdots \,\, \Jbf_D] \trans (\vect \Xbf),
\end{eqnarray*}
where $\Jbf_d \in \real{\prod_{d=1}^D p_d \times p_dR_d}$ is the Jacobian
\begin{eqnarray}
    \Jbf_d = D\Bbf(\Bbf_d) =  \Pibf_d \{ [(\Bbf_D \otimes \cdots \otimes \Bbf_{d+1} \otimes \Bbf_{d-1} \otimes \cdots \otimes \Bbf_1) \Gbf_{(d)} \trans] \otimes \Ibf_{p_d}\} \label{eqn:Jd}
\end{eqnarray}
and $\Pibf_d$ is the $(\prod_{d=1}^D p_d)$-by-$(\prod_{d=1}^D p_d)$ permutation matrix that reorders $\vect \Bbf_{(d)}$ to obtain $\vect \Bbf$, i.e., $
    \vect \Bbf = \Pibf_d \, \vect \Bbf_{(d)}.
$
\item Let the Hessian $d^2 \eta(\Gbf,\Bbf_1,\ldots,\Bbf_D) \in \real{(\prod_d R_d + \sum_d p_dR_d) \times (\prod_d R_d + \sum_d p_dR_d)}$ be partitioned into four blocks $\Hbf_{\Gbf,\Gbf} \in \real{\prod_d R_d \times \prod_d R_d}$, $\Hbf_{\Gbf,\Bbf} = \Hbf_{\Bbf,\Gbf} \trans \in \real{\prod_d R_d \times \sum_d p_d R_d}$ and $\Hbf_{\Bbf,\Bbf} \in \real{\sum_d p_d R_d \times \sum_d p_d R_d}$. Then $\Hbf_{\Gbf,\Gbf}={\bf 0}$, $\Hbf_{\Gbf,\Bbf}$ has entries
    \begin{eqnarray*}
        h_{(r_1,\ldots,r_D),(i_{d},s_{d})} &=&     1_{\{r_d=s_d\}} \sum_{j_d=i_d} x_{j_1,\ldots,j_D} \prod_{d'\ne d} \beta_{j_{d'}}^{(r_{d'})},
    \end{eqnarray*}
    and $\Hbf_{\Bbf,\Bbf}$ has entries
\begin{eqnarray*}
    h_{(i_d,r_d),(i_{d'},r_{d'})} = 1_{\{d\ne d'\}} \sum_{j_d=i_d,j_{d'}=i_{d'}} x_{j_1,\ldots,j_D} \sum_{s_d=r_d,s_{d'}=r_{d'}} g_{s_1,\ldots,s_D} \prod_{d''\ne d, d'} \beta_{j_{d''}}^{(s_{d''})}.
\end{eqnarray*}
Furthermore, $\Hbf_{\Bbf,\Bbf}$ can be partitioned in $D^2$ sub-blocks as
\begin{eqnarray*}
    \left( \begin{array}{cccc}
    {\bf 0} & * & * & * \\
    \Hbf_{21} & {\bf 0} & * & *  \\
    \vdots & \vdots & \ddots & *  \\
    \Hbf_{D1} & \Hbf_{D2} & \cdots & {\bf 0}
    \end{array}
    \right).
\end{eqnarray*}
The elements of sub-block $\Hbf_{dd'} \in \real{p_dR_d \times p_{d'}R_{d'}}$ can be retrieved from the matrix
$$
    \Xbf_{(dd')} (\Bbf_D \otimes \cdots \otimes \Bbf_{d+1} \otimes \Bbf_{d-1} \otimes \cdots \otimes \Bbf_{d'+1} \otimes \Bbf_{d'-1} \otimes \cdots \otimes \Bbf_1) \Gbf_{(dd')} \trans.
$$
$\Hbf_{\Gbf,\Bbf}$ can be partitioned into $D$ sub-blocks as $(\Hbf_1,\ldots,\Hbf_D)$. The sub-block $\Hbf_d \in \real{\prod_d R_d \times p_d R_d}$ has at most $p_d \prod_d R_d$ nonzero entries which can be retrieved from the matrix
\begin{eqnarray*}
    \Xbf_{(d)} (\Bbf_D \otimes \cdots \otimes \Bbf_{d+1} \otimes \Bbf_{d-1} \otimes \cdots \otimes \Bbf_1).
\end{eqnarray*}
\end{enumerate}
\end{lemma}

Let
$
    \ell(\Bbf_1,\ldots,\Bbf_D|y,\xbf) = \ln p (y|\xbf,\Bbf_1,\ldots,\Bbf_D)
$
be the log-density of GLM. Next result derives the score function, Hessian, and Fisher information of the Tucker tensor regression model.
\begin{proposition}
\label{prop:score-info}
Consider the tensor regression model defined by (\ref{eqn:GLM-density}) and (\ref{eqn:r-tensorreg}).
\begin{enumerate}
\item The score function (or score vector) is
\begin{align}
    \nabla \ell(\Gbf,\Bbf_1,\ldots,\Bbf_D) = \frac{(y - \mu) \mu'(\eta)}{\sigma^2} \nabla \eta(\Gbf,\Bbf_1,\ldots,\Bbf_D)    \label{eqn:tensor-score}
\end{align}
with $\nabla \eta(\Gbf,\Bbf_1,\ldots,\Bbf_D)$ given in Lemma \ref{lemma:eta-derivatives}.
\item The Hessian of the log-density $\ell$ is
\begin{eqnarray}
    & & H(\Gbf, \Bbf_1,\ldots,\Bbf_D) \nonumber \\
    &=& - \left[ \frac{[\mu'(\eta)]^2}{\sigma^2} - \frac{(y-\mu)\theta''(\eta)}{\sigma^2} \right] \nabla \eta(\Gbf,\Bbf_1,\ldots,\Bbf_D) d \eta(\Gbf,\Bbf_1,\ldots,\Bbf_D) \nonumber \\
    & & + \frac{(y-\mu)\theta'(\eta)}{\sigma^2} d^2 \eta(\Gbf, \Bbf_1,\ldots,\Bbf_D), \label{eqn:tensor-hessian}
\end{eqnarray}
with $d^2\eta$ defined in Lemma \ref{lemma:eta-derivatives}.
\item The Fisher information matrix is
\begin{eqnarray}
    & & \Ibf(\Gbf, \Bbf_1,\ldots,\Bbf_D) \nonumber \\
    &=& E[- H(\Gbf, \Bbf_1,\ldots,\Bbf_D)] \nonumber \\
    &=& \mathrm{Var} [ \nabla \ell(\Gbf, \Bbf_1,\ldots,\Bbf_D) d\ell(\Gbf, \Bbf_1,\ldots,\Bbf_D)] \nonumber \\
    &=& \frac{[\mu'(\eta)]^2}{\sigma^2}  [\Bbf_D \otimes \cdots \otimes \Bbf_1 \,\, \Jbf_1 \ldots \Jbf_D] \trans (\vect \Xbf) (\vect \Xbf) \trans [\Bbf_D \otimes \cdots \otimes \Bbf_1 \,\, \Jbf_1 \ldots \Jbf_D]. \label{eqn:fisher-info}
\end{eqnarray}
\end{enumerate}
\end{proposition}

\noindent
\textit{Remark 2.1: }
For canonical link, $\theta=\eta$, $\theta'(\eta)=1$, $\theta''(\eta)=0$, and the second term of Hessian vanishes. For the classical GLM with linear systematic part ($D=1$), $d^2 \eta(\Gbf,\Bbf_1,\ldots,\Bbf_D)$ is zero and  thus the third term of Hessian vanishes. For the classical GLM ($D=1$) with canonical link, both second and third terms of the Hessian vanish and thus the Hessian is non-stochastic, coinciding with the information matrix.

\subsection{Identifiability}

The Tucker decomposition \eqref{eqn:tucker} is unidentifiable due to the nonsingular transformation indeterminacy. That is
\begin{eqnarray*}
    \llbracket \Gbf; \Bbf_1,\ldots,\Bbf_D \rrbracket = \llbracket \Gbf \times_1 \Obf_1^{-1} \times \cdots \times_D \Obf_D^{-1}; \Bbf_1 \Obf_1,\ldots,\Bbf_D \Obf_D \rrbracket
\end{eqnarray*}
for any nonsingular matrices $\Obf_d \in \real{R_d \times R_d}$. This implies that the number of free parameters for a Tucker model is $\sum_d p_d R_d + \prod_d R_d - \sum_d R_d^2$, with the last term adjusting for nonsingular indeterminacy. Therefore the Tucker model is identifiable only in terms of the equivalency classes. 

For asymptotic consistency and normality, it is necessary to adopt a specific constrained parameterization. It is common to impose the orthonormality constraint on the factor matrices $\Bbf_d \trans \Bbf_d = \Ibf_{R_d}$, $d=1,\ldots,D$. However the resulting parameter space is a manifold and much harder to deal with. We adopt an alternative parameterization that fixes the entries of the first $R_d$ rows of $\Bbf_d$ to be ones
\begin{eqnarray*}
    {\cal \Bbf} = \{\llbracket \Gbf; \Bbf_1,\ldots,\Bbf_D \rrbracket: \beta_{i_d}^{(r)} = 1, i_d=1,\ldots,R_d, d=1,\ldots,D \}.
\end{eqnarray*}
The formulae for score, Hessian and information in Proposition \ref{prop:score-info} require changes accordingly. The entries in the first $R_d$ rows of $\Bbf_d$ are fixed at ones and their corresponding entries, rows and columns in score, Hessian and information need to be deleted. Choice of the restricted space $\mathcal{\Bbf}$ is obviously arbitrary, and excludes arrays with any entries in the first rows of $\Bbf_d$ equal to zeros. However the set of such exceptional arrays has Lebesgue measure zero. In specific applications, subject knowledge may suggest alternative restrictions on the parameters.

Given a finite sample size, conditions for global identifiability of parameters are in general hard to obtain except in the linear case ($D=1$). 
Local identifiability essentially requires linear independence between the ``collapsed" vectors $[\Bbf_D \otimes \cdots \otimes \Bbf_1 \,\, \Jbf_1 \ldots \Jbf_D] \trans \vect \xbf_i \in \real{\sum_d p_d R_d + \prod_d R_d - \sum_d R_d^2}$.
\begin{proposition}[Identifiability]
\label{prop:identifiability}
Given iid data points $\{(y_i,\xbf_i),i=1,\ldots,n\}$ from the Tucker tensor regression model. Let $\Bbf_0 \in \mathcal{\Bbf}$ be a parameter point and assume there exists an open neighborhood of $\Bbf_0$ in which the information matrix has a constant rank. Then $\Bbf_0$ is locally identifiable if and only if
\begin{align*}
    I(\Bbf_0) = [\Bbf_D \otimes \cdots \otimes \Bbf_1 \,\, \Jbf_1 \ldots \Jbf_D] \trans \left[ \sum_{i=1}^n \frac{\mu'(\eta_i)^2}{\sigma_i^2} (\vect \, \xbf_i) (\vect \, \xbf_i) \trans \right] [\Bbf_D \otimes \cdots \otimes \Bbf_1 \,\, \Jbf_1 \ldots \Jbf_D]
\end{align*}
is nonsingular.
\end{proposition}

\subsection{Asymptotics}

The asymptotics for tensor regression follow from those for MLE or M-estimation. The key observation is that the nonlinear part of tensor model (\ref{eqn:r-tensorreg}) is a degree-$D$ polynomial of parameters and the collection of polynomials $\{\langle \Bbf, \Xbf \rangle, \Bbf \in \mathcal{\Bbf}\}$ form a Vapnik-\u{C}ervonenkis (VC) class. Then the classical uniform convergence theory applies \citep{vanderVaart98Asymp}. For asymptotic normality, we need to establish that the log-likelihood function of tensor regression model is quadratic mean differentiable \citep{lehmannRomano05TSH}. A sketch of the proof is given in the Appendix.
\begin{theorem}
\label{thm:consistency}
Assume $\Bbf_0 \in \mathcal{\Bbf}$ is (globally) identifiable up to permutation and  the array covariates $\Xbf_i$ are iid from a bounded underlying distribution.
\begin{enumerate}
\item (Consistency) The MLE is consistent, i.e., $\hat \Bbf_n$ converges to $\Bbf_0$ in probability, in following models. (1) Normal tensor regression with a compact parameter space $\mathcal{\Bbf}_0 \subset \mathcal{\Bbf}$. (2) Binary tensor regression. (3) Poisson tensor regression with a compact parameter space $\mathcal{\Bbf}_0 \subset \mathcal{\Bbf}$.
\item (Asymptotic Normality) For an interior point $\Bbf_0 \in \mathcal{\Bbf}$ with nonsingular information matrix $\Ibf(\Bbf_0)$ (\ref{eqn:fisher-info}) and $\hat \Bbf_n$ is consistent, $\sqrt n (\vect \hat \Bbf_n - \vect \Bbf_0)$ converges in distribution to a normal with mean zero and covariance matrix $\Ibf^{-1}(\Bbf_0)$.
\end{enumerate}
\end{theorem}

In practice it is rare that the true regression coefficient $\Bbf_{\text{true}} \in \real{p_1 \times \cdots \times p_D}$ is exactly a low rank tensor. However the MLE of the rank-$R$ tensor model converges to the maximizer of function $M(\Bbf) = \mathbb{P}_{\Bbf_{\text{true}}} \ln p_{\Bbf}$ or equivalently $\mathbb{P}_{\Bbf_{\text{true}}} \ln (p_{\Bbf}/p_{\Bbf_{\text{true}}})$. In other words, the MLE consistently estimates the best approximation (among models in ${\cal \Bbf}$) of $\Bbf_{\text{true}}$ in the sense of Kullback-Leibler distance.

\section{Regularized Estimation}
\label{sec:regularization}

Regularization plays a crucial role in neuroimaging analysis for several reasons. First, even after substantial dimension reduction by imposing a Tucker structure, the number of parameters $p_{\text{T}}$ can still exceed the number of observations $n$. Second, even when $n > p_{\text{T}}$, regularization could potentially be useful for stabilizing the estimates and improving the risk property. Finally, regularization is an effective way to incorporate prior scientific knowledge about brain structures. For instance, it may sometimes be reasonable to impose symmetry on the parameters along the coronal plane for MRI images. 

In our context of Tucker regularized regression, there are two possible types of regularizations, one on the core tensor $\Gbf$ \emph{only}, and the other on both $\Gbf$ and $\Bbf_d$ \emph{simultaneously}. Which regularization to use depends on the practical purpose of a scientific study. In this section, we illustrate the regularization on the core tensor, which simultaneously achieves sparsity in the number of outer products in Tucker decomposition \eqref{eqn:tucker} and shrinkage. Toward that purpose, we propose to maximize the regularized log-likelihood
\begin{eqnarray*}
    \ell(\gammabf, \Gbf, \Bbf_1, \ldots,\Bbf_D) - \sum_{r_1,\ldots,r_D} P_\eta(|g_{r_1,\ldots,r_D}|,\lambda),
\end{eqnarray*}
where $P_\eta(|x|,\lambda)$ is a scalar penalty function, $\lambda$ is the penalty tuning parameter, and $\eta$ is an index for the penalty family. Note that the penalty term above only involves elements of the core tensor, and thus regularization on $\Gbf$ only. This formulation includes a large class of penalty functions, including power family \citep{FrankFriedman93Bridge},  where $P_\eta(|x|,\lambda) = \lambda |x|^\eta$, $\eta \in (0,2]$, and in particular lasso \citep{Tibshirani96Lasso} ($\eta=1$) and ridge ($\eta=2$); elastic net \citep{ZouHastie05Enet}, where $P_\eta(|x|, \lambda) = \lambda [(\eta-1) x^2/2 + (2-\eta) |x|]$, $\eta \in [1,2]$; SCAD \citep{FanLi01SCAD}, where $\partial / \partial |x| P_{\eta}(|x|, \lambda) = \lambda \left\{ 1_{\{|x| \le \lambda\}} + (\eta \lambda - |x|)_+ /(\eta-1)\lambda 1_{\{|x| > \lambda\}} \right\}$, $\eta>2$; and MC+ penalty \citep{Zhang10MCP}, where $P_{\eta}(|x|, \lambda) = \left\{\lambda|x|-x^2 / (2\eta) \right\} 1_{\{|x|<\eta \lambda\}} + 0.5 \lambda^2 \eta 1_{\{|x| \ge \eta \lambda\}}$, among many others.

Two aspects of the proposed regularized Tucker regression, parameter estimation and tuning, deserve some discussion. For regularized estimation, it incurs only slight changes in Algorithm \ref{algo:br-algo}. That is, when updating $\Gbf$, we simply fit a penalized GLM regression problem,
\begin{align*}
    \Gbf^{(t+1)} = \mbox{argmax}_{\Gbf} \, \ell(\gammabf^{(t)}, \Bbf_1^{(t+1)}, \ldots, \Bbf_D^{(t+1)}, \Gbf) - \sum_{r_1,\ldots,r_D} P_\eta(|g_{r_1,\ldots,r_D}|,\lambda),
\end{align*}
for which many software packages exist. Our implementation utilizes an efficient {\sc Matlab} toolbox for sparse regression \citep{ZhouArmagan11SparsePath}. Other steps of Algorithm \ref{algo:br-algo} remain unchanged. For the regularization to remain legitimate, we constrain the column norms of $\Bbf_d$ to be one when updating factor matrices $\Bbf_d$. For parameter tuning, one can either use the general cross validation approach, or employ Bayesian information criterion to tune the penalty parameter $\lambda$. 

\section{Numerical Study}
\label{sec:numerics}

We have carried out intensive numerical experiments to study the finite sample performance of the Tucker regression. Our simulations focus on three aspects: first, we demonstrate the capacity of the Tucker regression in identifying various shapes of signals; second, we study the consistency property of the method by gradually increasing the sample size; third, we compare the performance of the Tucker regression with the CP regression of \citet{ZhouLiZhu12CPTensor}. We also examine a real MRI imaging data to illustrate the Tucker downsizing and to further compare the two tensor models.

\subsection{Identification of Various Shapes of Signals}
\label{sec:shapes}

\begin{figure}
\begin{center}
\begin{tabular}{c}
\includegraphics[width=4.4in]{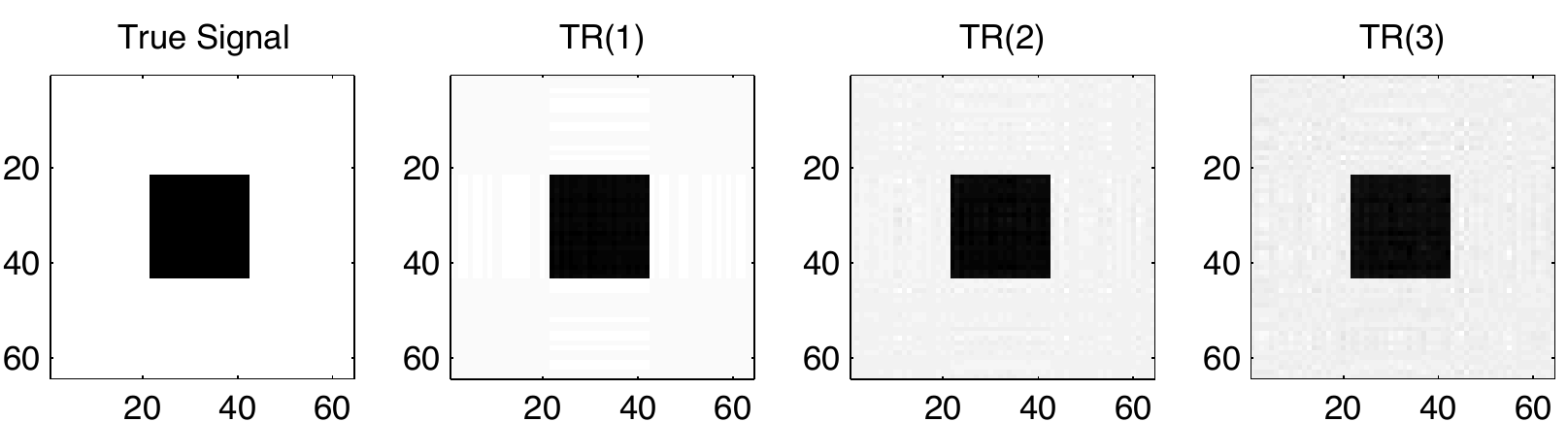} \\ \includegraphics[width=4.4in]{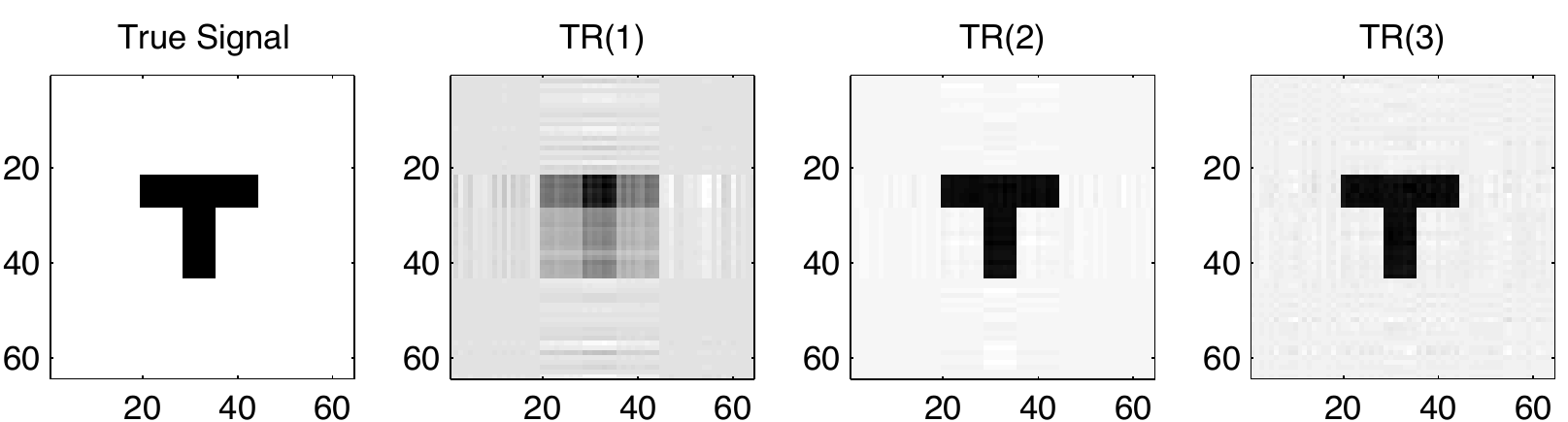}    \\
\includegraphics[width=4.4in]{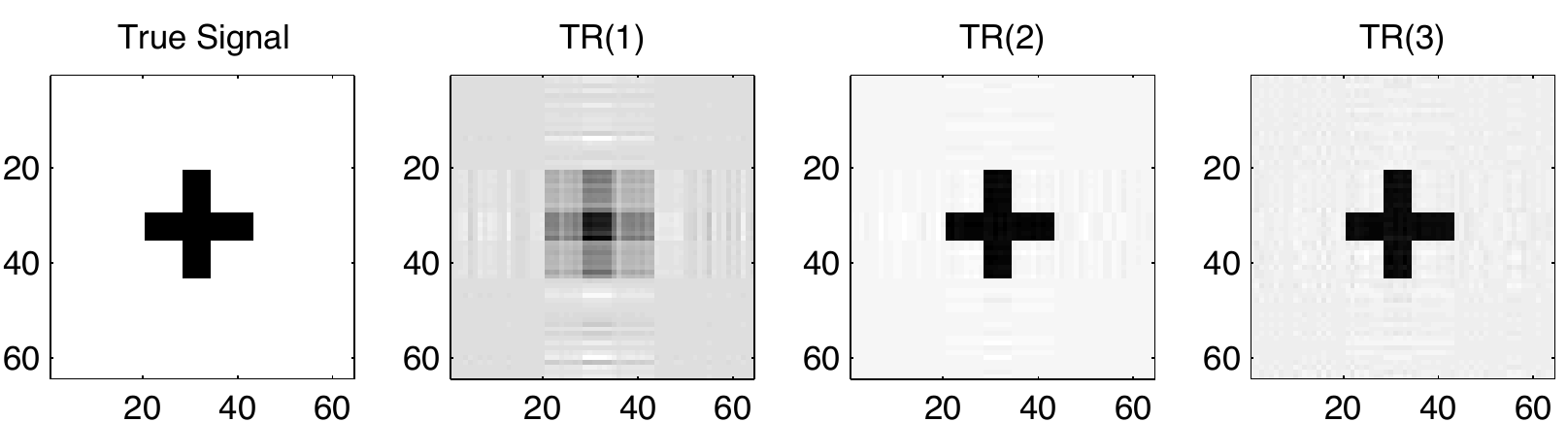} \\ \includegraphics[width=4.4in]{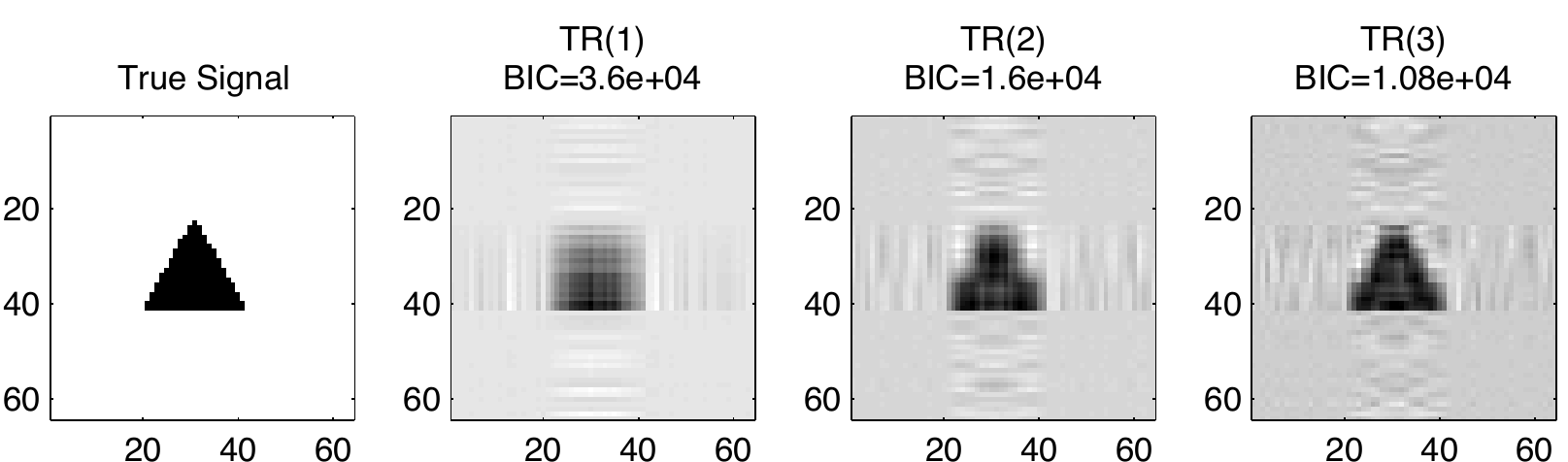}    \\
\includegraphics[width=4.4in]{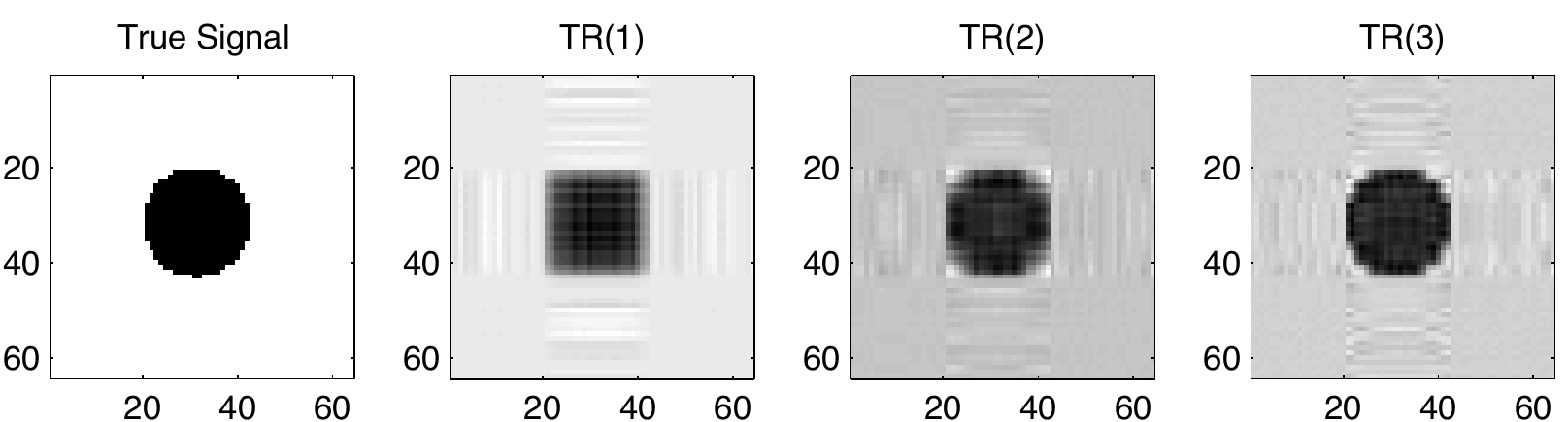} \\ \includegraphics[width=4.4in]{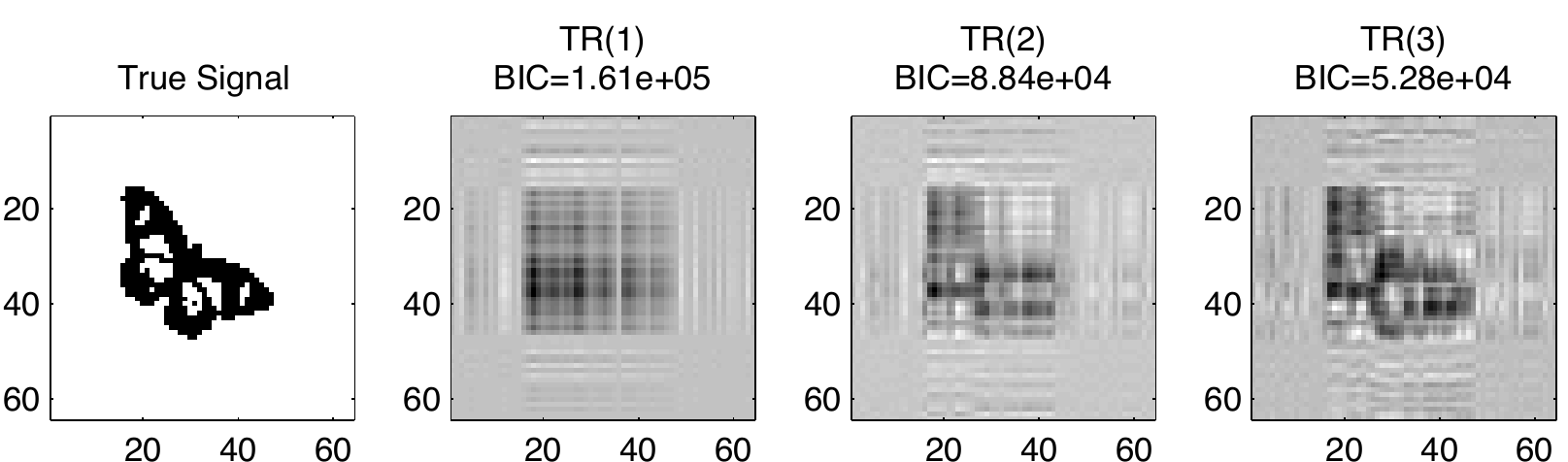}
\end{tabular}
\caption{True and recovered image signals by Tucker regression. The matrix variate has size 64 by 64 with entries generated as independent standard normals. The regression coefficient for each entry is either 0 (white) or 1 (black). The sample size is 1000. TR$(r)$ means estimate from the Tucker regression with an $r$-by-$r$ core tensor.}
\label{fig:shapes}
\end{center}
\end{figure}

In our first example, we demonstrate that the proposed Tucker regression model, though with substantial reduction in dimension, can manage to identify a range of two dimensional signal shapes with varying ranks. In Figure \ref{fig:shapes}, we list the 2D signals $\Bbf \in \real{64 \times 64}$ in the first row, along with the estimates by Tucker tensor models in the second to fourth rows with orders $(1,1), (2,2)$ and $(3,3)$, respectively. Note that, since the orders along both dimensions are made equal, the Tucker model is to perform essentially the same as a CP model in this example, and the results are presented here for completeness. We will examine differences of the two models in later examples.  The regular covariate vector $\Zbf \in \real{5}$ and image covariate $\Xbf \in \real{64 \times 64}$ are randomly generated with all elements being independent standard normals. The response $Y$ is generated from a normal model with mean $\mu = \gammabf \trans \Zbf + \langle \Bbf, \Xbf \rangle$ and variance $\textrm{var}(\mu)/10$. The vector coefficient $\gammabf = {\bf 1}_5$, and the coefficient array $\Bbf$ is binary, with the signal region equal to one and the rest zero. Note that this problem differs from the usual edge detection or object recognition in imaging processing \citep{Qiu2005book,Qiu07jumpsurface}. In our setup, all elements of the image $\Xbf$ follow the same distribution. The signal region is defined through the coefficient matrix $\Bbf$ and needs to be inferred from the relation between $Y$ and $\Xbf$ after adjusting for $\Zbf$. It is clearly see in Figure \ref{fig:shapes} that, the Tucker model yields a sound recovery of the true signals, even for those of high rank or natural shape, e.g., ``disk" and ``butterfly". We also illustrate in the plot the BIC criterion in Section ~\ref{sec:tucker-vs-cp}. 

\subsection{Performance with Increasing Sample Size}

\begin{figure}
\begin{center}
\begin{tabular}{cc}
{\centering $p_1 = p_2 = p_3 = 16$} & {\centering $p_1 = p_2 = p_3 = 32$} \\
\includegraphics[width=2.4in]{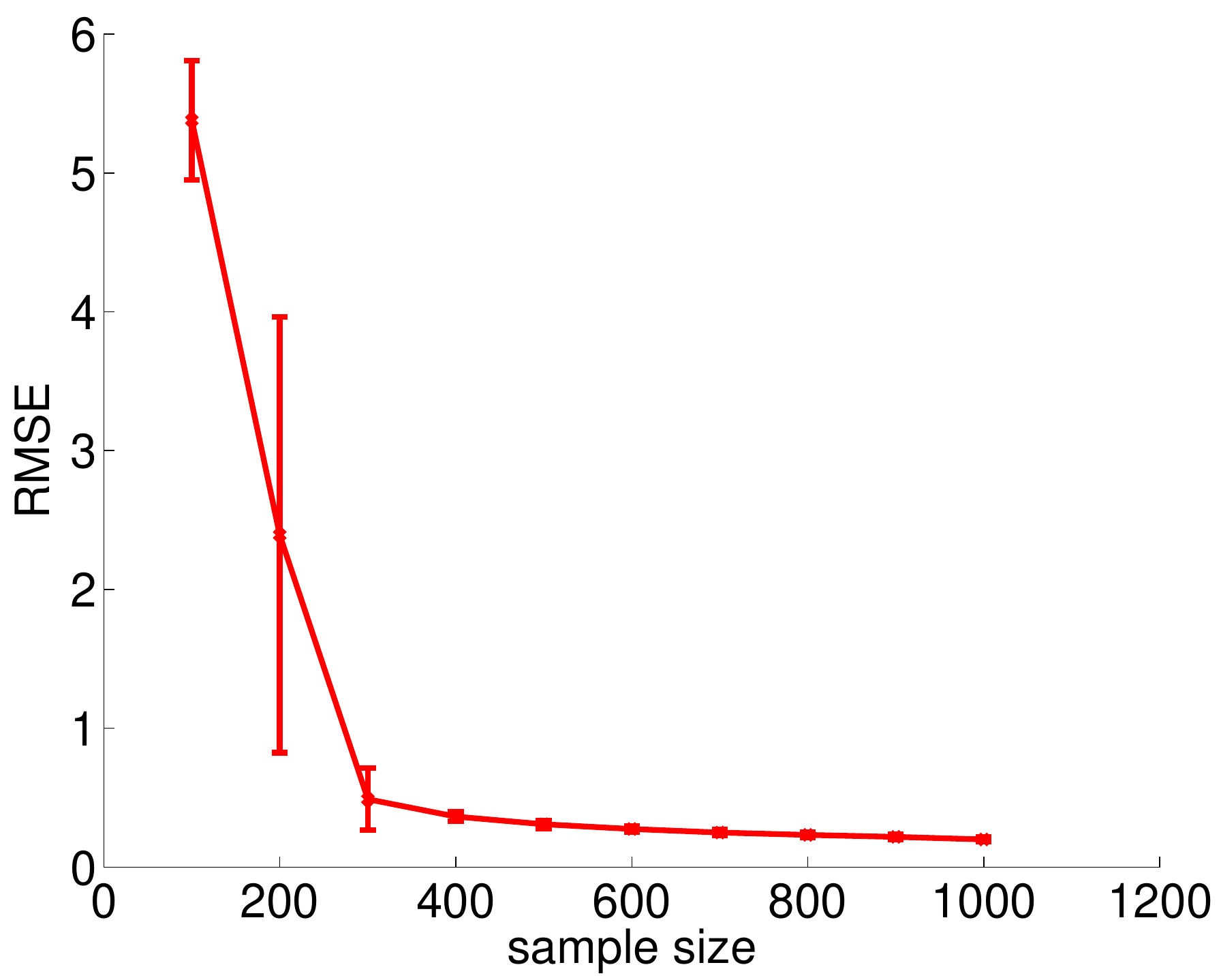} & \includegraphics[width=2.4in]{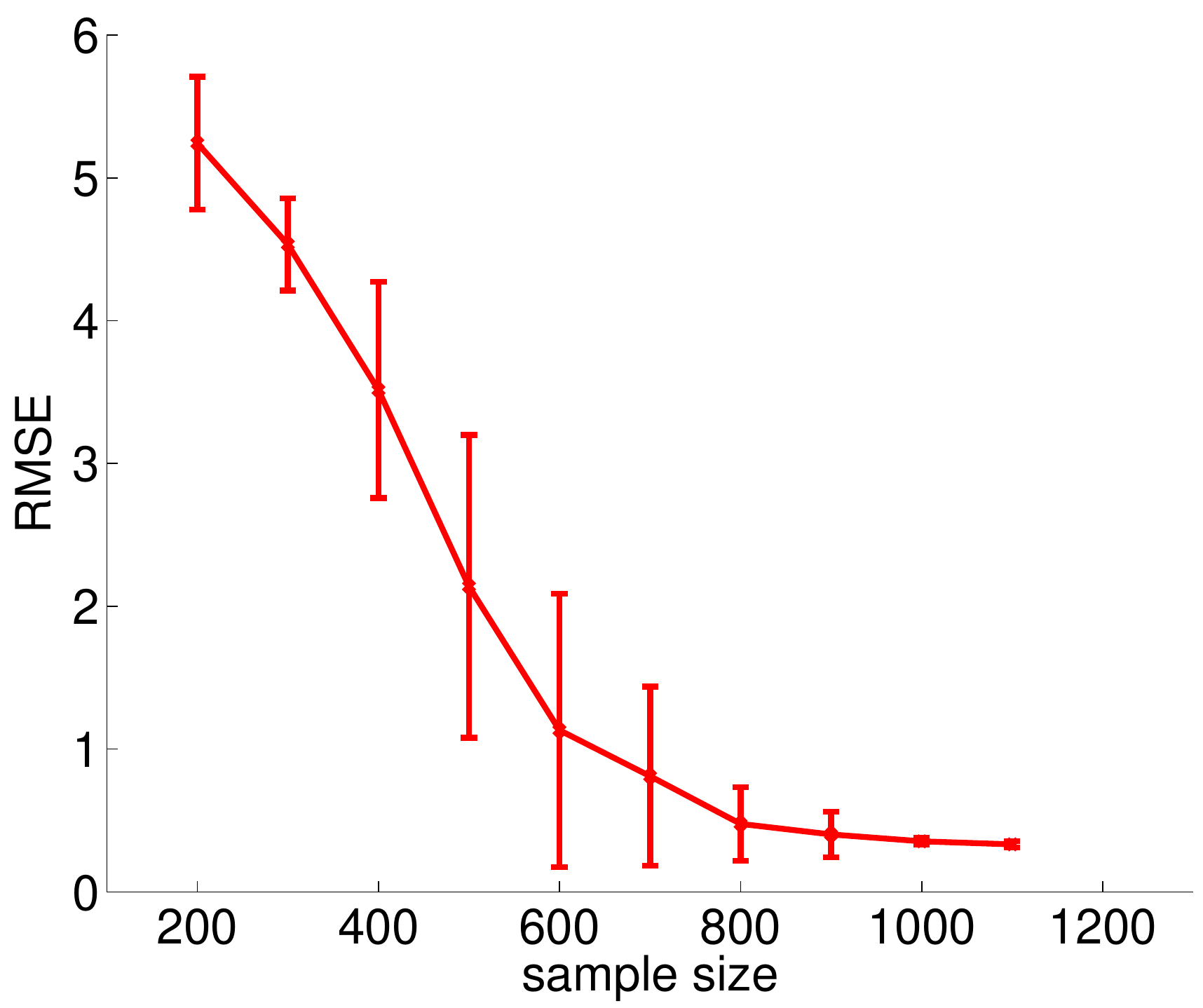}    \\
\includegraphics[width=2.4in]{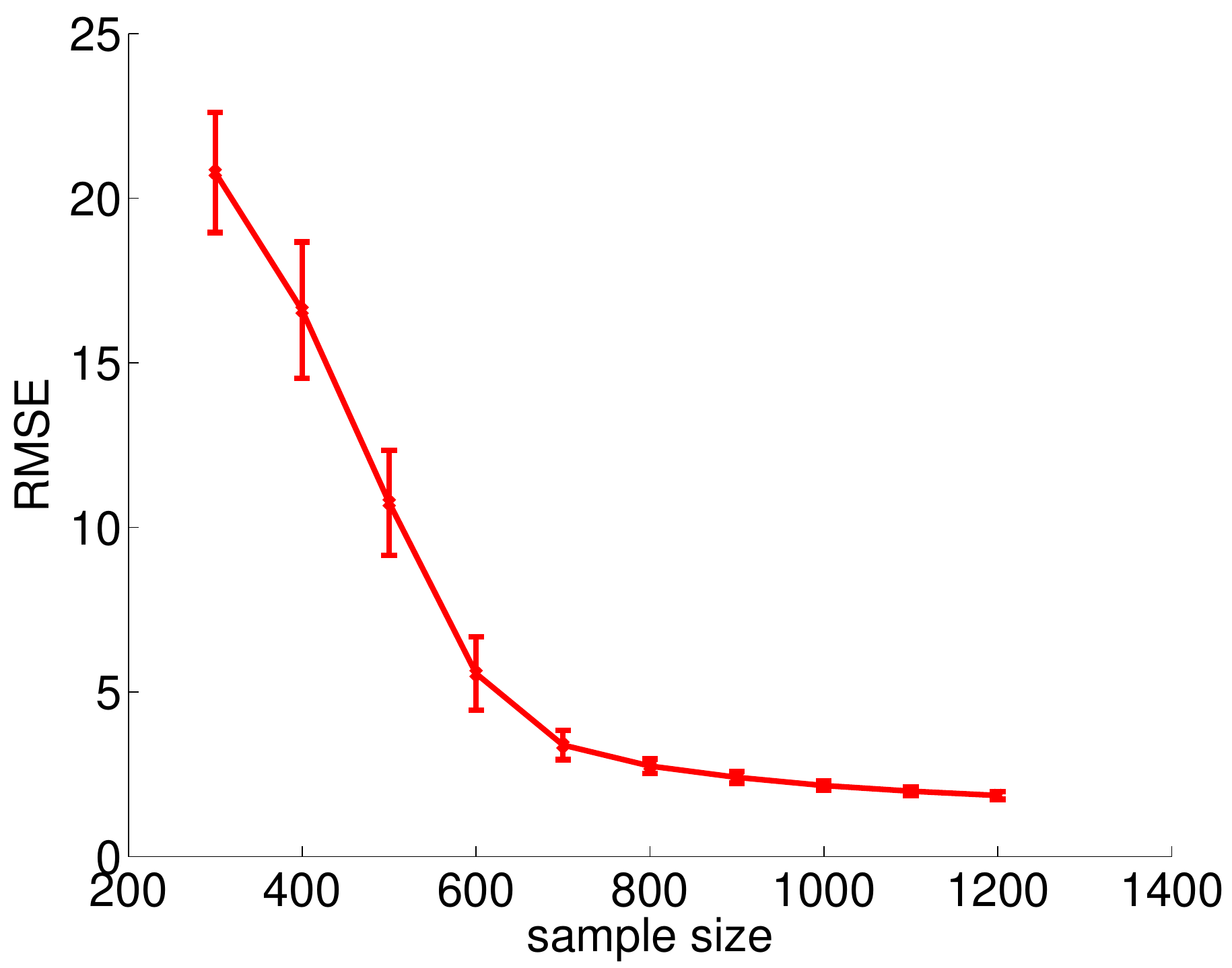} & \includegraphics[width=2.4in]{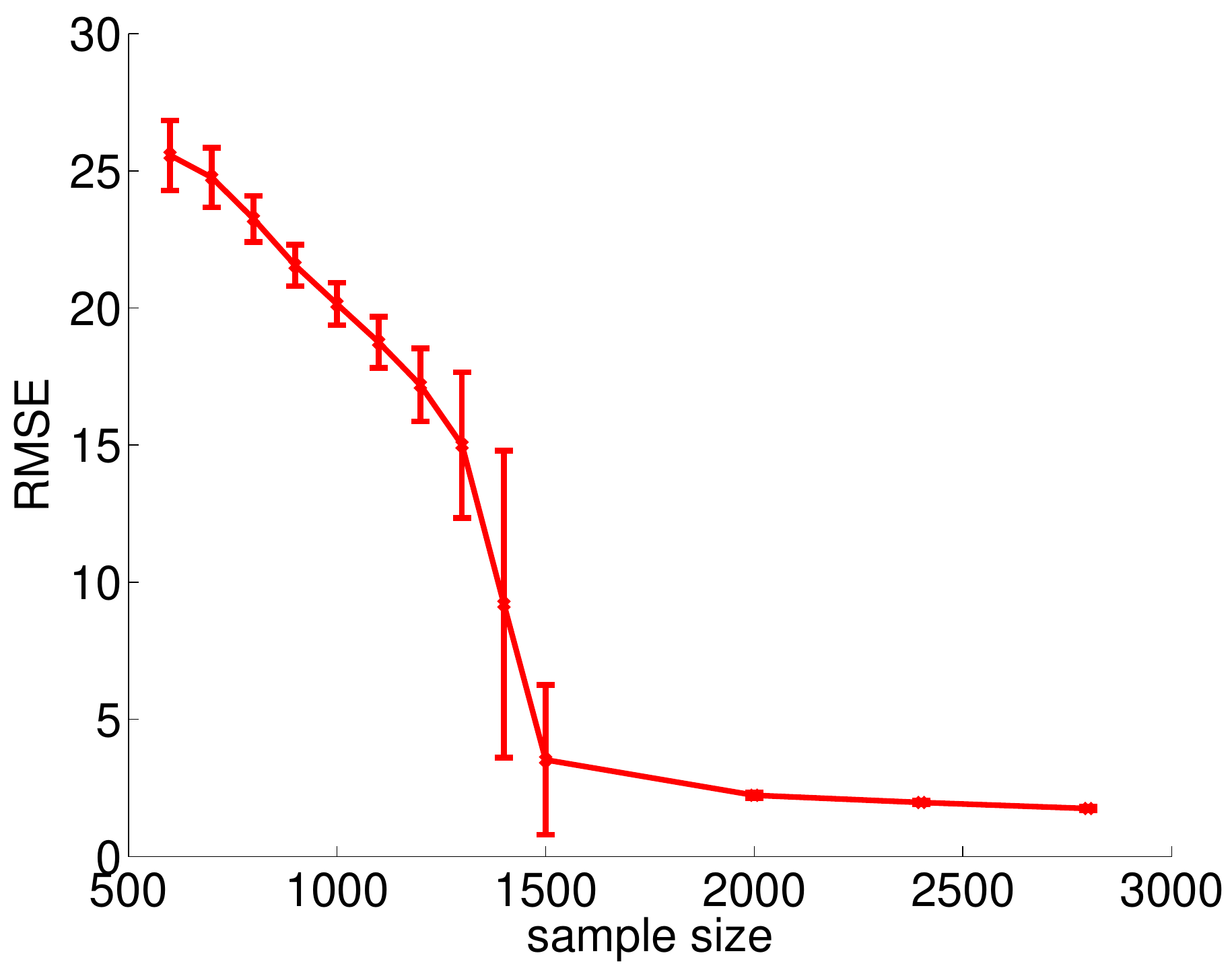}    \\
\includegraphics[width=2.4in]{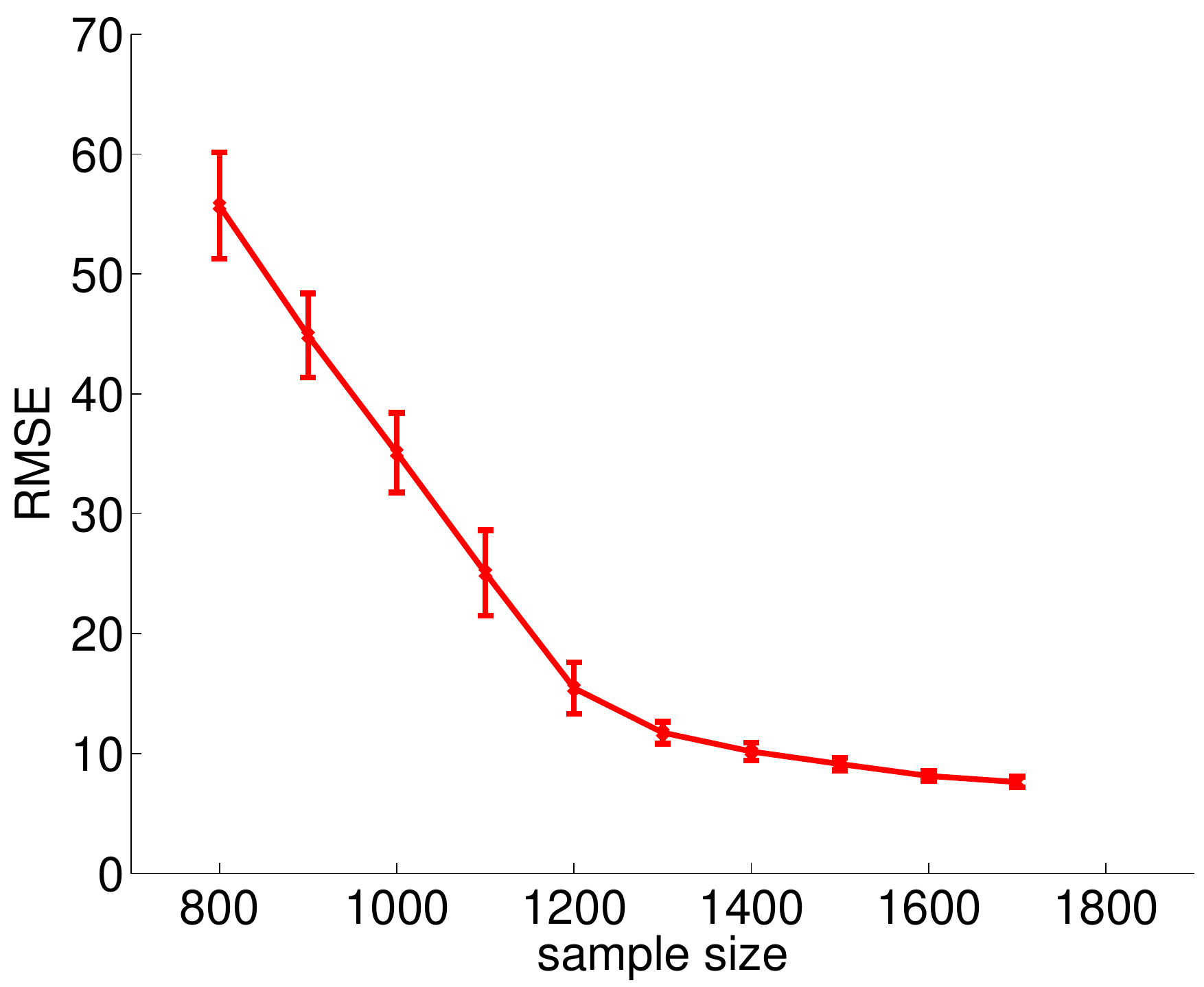} & \includegraphics[width=2.4in]{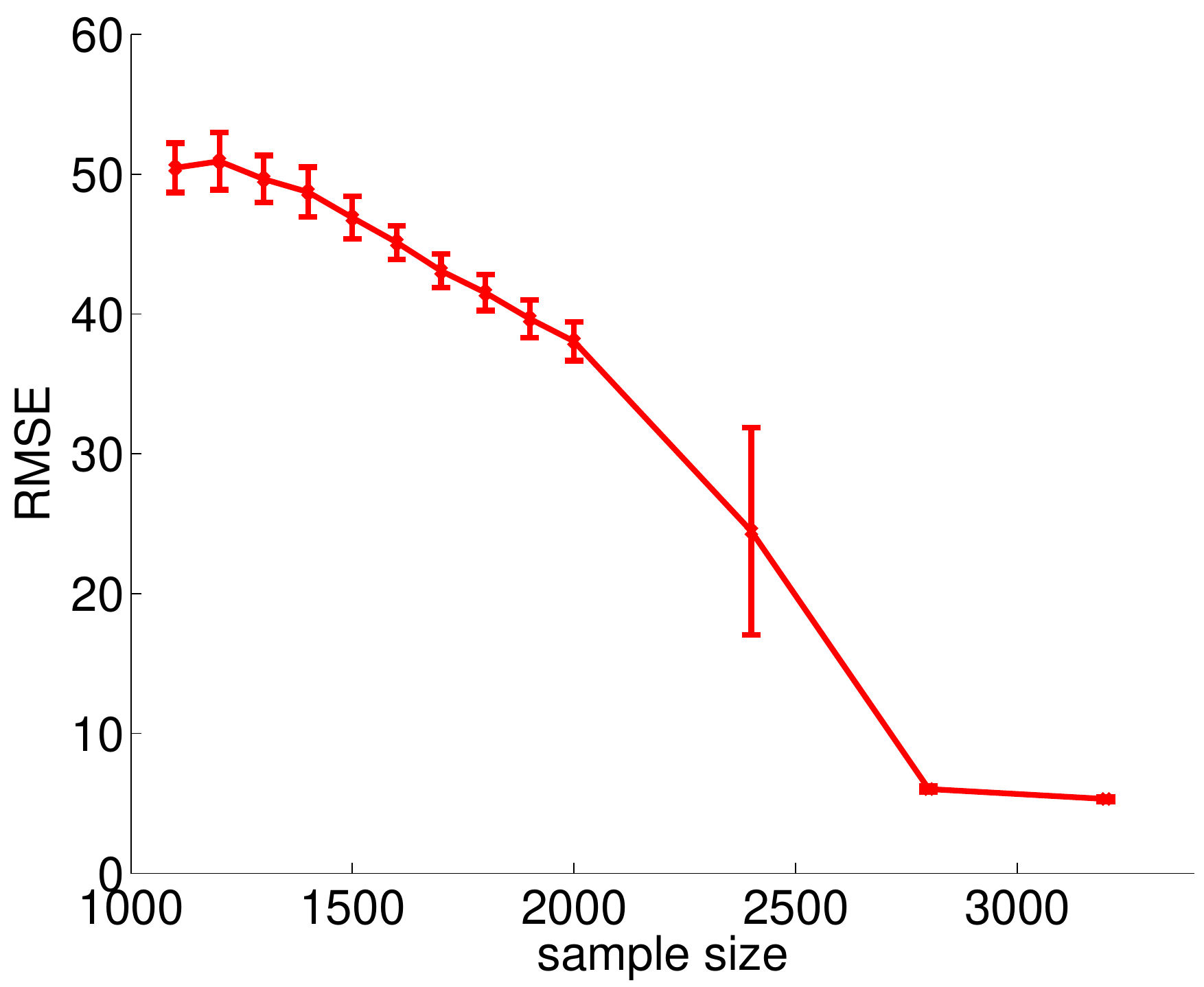}
\end{tabular}
\caption{Root mean squared error (RMSE) of the tensor parameter estimate versus the sample size. Reported are the average and standard deviation of RMSE based on 100 data replications. Top: $R_1=R_2=R_3=2$; Middle: $R_1=R_2=R_3=5$; Bottom: $R_1=R_2=R_3=8$.}
\label{fig:size}
\end{center}
\end{figure}

In our second example, we continue to employ a similar model as in Figure \ref{fig:shapes} but with a three dimensional image covariate. The dimension of $\Xbf$ is set as $p_1\times p_2 \times p_3$, with $p_1 = p_2 = p_3 = 16$ and $32$, respectively. The signal array $\Bbf$ is generated from a Tucker structure, with the elements of core tensor $\Gbf$ and the factor matrices $\Bbf$'s all coming from independent standard normals. The dimension of the core tensor $\Gbf$ is set as $R_1 \times R_2 \times R_3$, with $R_1 = R_2 = R_3 = 2, 5$, and $8$, respectively. We gradually increase the sample size, starting with an $n$ that is in hundred and no smaller than the degrees of freedom of the generating model. We aim to achieve two purposes with this example: first, we verify the consistency property of the proposed estimator, and second, we gain some practical knowledge about the estimation accuracy with different values of the sample size. Figure \ref{fig:size} summarizes the results. It is clearly seen that the estimation improves with the increasing sample size. Meanwhile, we observe that, unless the core tensor dimension is small, one would require a relatively large sample size to achieve a good estimation accuracy. This is not surprising though, considering the number of parameters of the model and that regularization is not employed here. The proposed tensor regression approach has been primarily designed for imaging studies with a reasonably large number of subjects. Recently, a number of such large-scale brain imaging studies are emerging. For instance, the Attention Deficit Hyperactivity Disorder Sample Initiative \citep{ADHD-url} consists of over 900 participants from eight imaging centers with both MRI and fMRI images, as well as their clinical information. Another example is the Alzheimer's Disease Neuroimaging Initiative \citep{ADNI-url} database, which accumulates over 3,000 participants with MRI, fMRI and genomics data. In addition, regularization discussed in Section \ref{sec:regularization} and the Tucker downsizing in Section \ref{sec:duality} can both help improve estimation given a limited sample size. 

\subsection{Comparison of the Tucker and CP Models}
\label{sec:num-compare}

In our third example, we focus on comparison between the Tucker tensor model with the CP tensor model of \citet{ZhouLiZhu12CPTensor}. 
We generate a normal response, and the 3D signal array $\Bbf$ with dimensions $p_1, p_2, p_3$ and the $d$-ranks $r_1, r_2, r_3$.  Here, the $d$-rank is defined as the column rank of the mode-$d$ matricization $\Bbf_{(d)}$ of $\Bbf$. We set $p_1 = p_2 = p_3 = 16$ and $32$, and $(r_1, r_2, r_3) = (5,3,3), (8,4,4)$ and $(10,5,5)$, respectively. The sample size is 2000. We fit a Tucker model with $R_d = r_d$, and a CP model with $R = \max r_d$, $d = 1, 2, 3$. We report in Table ~\ref{tab:compare} the degrees of freedom of the two models under different setup, as well as the root mean squared error (RMSE) out of 100 data replications. It is seen that the Tucker model requires a smaller number of free parameters, while it achieves a more accurate estimation compared to the CP model. Such advantages come from the flexibility of the Tucker decomposition that permits different orders $R_d$ along directions. 

\begin{table}[t]
\caption{Comparison of the Tucker and CP models. Reported are the average and standard deviation (in the parenthesis) of the root mean squared error, all based on 100 data replications.}
\label{tab:compare}
\vspace{-0.05in}
\begin{center}
\begin{tabular}{|lll|ccc|} \hline
Dimension & Criterion & Model  & $(5,3,3)$ & $(8,4,4)$ & $(10,5,5)$ \\ \hline
$16\times16\times16$ & Df        & Tucker & 178 & 288 & 420 \\
                                         &             & CP        & 230 & 368 & 460 \\
                                         & RMSE & Tucker & 0.202 (0.013) & 0.379 (0.017) & 0.728 (0.030) \\
                                         &              & CP       & 0.287 (0.033) & 1.030 (0.081) & 2.858 (0.133) \\ \hline
$32\times32\times32$ & Df         & Tucker & 354 & 544 & 740 \\
                                         &              & CP        & 470 & 752 & 940 \\                             
                                         & RMSE & Tucker & 0.288 (0.013) & 0.570 (0.023) & 1.236 (0.045) \\
                                         &              & CP        & 0.392 (0.046) & 1.927 (0.172) & 16.238 (3.867)\\ \hline
\end{tabular}                     
\end{center}
\end{table}

\subsection{Attention Deficit Hyperactivity Disorder Data Analysis}
\label{sec:real-data}

We analyze the attention deficit hyperactivity disorder (ADHD) data from the ADHD-200 Sample Initiative \citep{ADHD-url} to illustrate our proposed method as well as the Tucker downsizing. ADHD is a common childhood disorder and can continue through adolescence and adulthood. Symptoms include difficulty in staying focused and paying attention, difficulty in controlling behavior, and over-activity. The data set that we analyzed is part of the ADHD-200 Global Competition data sets. It was pre-partitioned into a training data of 770 subjects and a testing data of 197 subjects. We removed those subjects with missing observations or poor image quality, resulting in 762 training subjects and 169 testing subjects. In the training set, there were 280 combined ADHD subjects,  482 normal controls, and the case-control ratio is about 3:5. In the testing set, there were 76 combined ADHD subjects, 93 normal controls, and the case-control ratio is about 4:5. T1-weighted images were acquired for each subject, and were preprocessed by standard steps. The data we used is obtained from the Neuro Bureau after preprocessing (the Burner data, \textsf{http://neurobureau.projects.nitrc.org/ADHD200/Data.html}). In addition to the MRI image predictor, we also include the subjects' age and handiness as regular covariates. The response is the binary diagnosis status. 

The original image size was $p_1 \times p_2 \times p_3 = 121 \times 145 \times 121$. We employ the Tucker downsizing in Section \ref{sec:duality}. More specifically, we first choose a wavelet basis for $\Bbf_d \in \real{p_d \times \tilde p_d}$, then transform the image predictor from $\Xbf$ to $\tilde \Xbf = \llbracket \Xbf; \Bbf_1 \trans, \ldots, \Bbf_D \trans \rrbracket$. We pre-specify the values of $\tilde p_d$'s that are about tenth of the original dimensions $p_d$, and equivalently, we fit a Tucker tensor regression with the image predictor dimension downsized to $\tilde p_1 \times \tilde p_2 \times \tilde p_3$. In our example, we have experimented with a set of values of $\tilde p_d$'s, and the results are qualitatively similar. We report two sets, $\tilde p_1 = 12$, $\tilde p_2 = 14$, $\tilde p_3 = 12$, and $\tilde p_1 = 10$, $\tilde p_2 = 12$, $\tilde p_3 = 10$. We have also experimented with the Haar wavelet basis (Daubechies D2) and the Daubechies D4 wavelet basis, which again show similar qualitative patterns. 

For $\tilde p_1 = 12, \tilde p_2 = 14, \tilde p_3 = 12$, we fit a Tucker tensor model with $R_1 = R_2 = R_3 = 3$, resulting in 114 free parameters, and fit a CP tensor model with $R = 4$, resulting in 144 free parameters. For $\tilde p_1 = 10, \tilde p_2 = 12, \tilde p_3 = 10$, we fit a Tucker tensor model with $R_1 = R_2 = 2$ and $R_3 = 3$, resulting in 71 free parameters, and fit a CP tensor model with $R = 4$, resulting in 120 free parameters. We have chosen those orders based on the following considerations. First, the number of free parameters of the Tucker and CP models are comparable. Second, at each step of GLM model fit, we ensure that the ratio between the sample size $n$ and the number of parameters under estimation in that step $\tilde p_d \times R_d$ satisfies a heuristic rule of greater than two in normal models and greater than five in logistic models. In the Tucker model, we also ensure the ratio between $n$ and the number of parameters in the core tensor estimation $\prod_d R_d$ satisfies this rule. We note that this selection of Tucker orders is heuristic; however, it seems to be a useful guideline especially when the data is noisy. We also fit a regularized Tucker model and a regularized CP model with the same orders, while the penalty parameter is tuned based on 5-fold cross validation of the training data. 

We evaluate each model by comparing the misclassification error rate on the independent testing set. The results are shown in Table \ref{tab:adhd-error}. We see from the table that, the regularized Tucker model performs the best, which echoes the findings in our simulations above. We also remark that, considering the fact that the ratio of case-control is about 4:5 in the testing data, the misclassification rate from 0.32 to 0.36 achieved by the regularized Tucker model indicates a fairly sound classification accuracy. On the other hand, we note that, a key advantage of our proposed approach is its capability of suggesting a useful model rather than the classification accuracy per se. This is different from black-box type machine learning based imaging classifiers.

\begin{table}[t]
\caption{ADHD testing data misclassification error.}
\label{tab:adhd-error}
\vspace{-0.05in}
\begin{center}
\begin{tabular}{|c|c|cccc|} \hline
Basis & Reduced dimension & Reg-Tucker & Reg-CP & Tucker & CP \\ \hline
Haar (D2)  & $12 \times 14 \times 12$ & \mbox{ } 0.361 \mbox{ } & \mbox{ } 0.367 \mbox{ } & \mbox{ } 0.379 \mbox{ } & \mbox{ } 0.438 \mbox{ }  \\
            & $10 \times 12 \times 10$ & 0.343 & 0.390 & 0.379 & 0.408  \\ \hline
Daubechies (D4)  & $12 \times 14 \times 12$ & 0.337 & 0.385 & 0.385 & 0.414  \\
                        & $10 \times 12 \times 10$ & 0.320 & 0.396 & 0.367 & 0.373 \\ \hline
\end{tabular}
\end{center}
\end{table}

It is also of interest to compare the run times of the two tensor model fittings. We record the run times of fitting the Tucker and CP models with the ADHD training data in Table \ref{tab:runtime}. They are comparable.

\begin{table}[t]
\caption{ADHD model fitting run time (in seconds).}
\label{tab:runtime}
\vspace{-0.05in}
\begin{center}
\begin{tabular}{|c|c|cccc|} \hline
Basis & Reduced dimension & Reg-Tucker & Reg-CP & Tucker & CP \\ \hline
Haar (D2)   & $12 \times 14 \times 12$ & \mbox{ } 3.68 \mbox{ } & \mbox{ } 4.39 \mbox{ } & \mbox{ } 31.25 \mbox{ } & \mbox{ } 22.43 \mbox{ } \\
            & $10 \times 12 \times 10$ & 1.36 & 2.79 & 9.08 & 25.10\\ \hline
Daubechies (D4)  & $12 \times 14 \times 12$ & 3.30 & 2.18 & 16.87 & 26.34 \\
                        & $10 \times 12 \times 10$ & 1.92 & 1.90 & 9.96& 17.10 \\ \hline
\end{tabular}
\end{center}
\end{table}

\section{Discussion}
\label{sec:discussion}

We have proposed a tensor regression model based on the Tucker decomposition. Including the CP tensor regression  \citep{ZhouLiZhu12CPTensor} as a special case, Tucker model provides a more flexible framework for regression with imaging covariates. We develop a fast estimation algorithm, a general regularization procedure, and the associated asymptotic properties. In addition, we provide a detailed comparison, both analytically and numerically, of the Tucker and CP tensor models. 

In real imaging analysis, the signal hardly has an exact low rank. On the other hand, given the limited sample size, a low rank estimate often provides a reasonable approximation to the true signal. This is why the low rank models such as the Tucker and CP could offer a sound recovery of even a complex signal. 

The tensor regression framework established in this article is general enough to encompass a large number of potential extensions, including but not limited to imaging multi-modality analysis, imaging classification, and longitudinal imaging analysis. These extensions consist of our future research.


\bibliography{ref_tucker}

\begin{thebibliography}{}

\bibitem[ADHD, 2013]{ADHD-url}
ADHD (2013).
\newblock The {ADHD}-200 sample.
\newblock \url{http://fcon_1000.projects.nitrc.org/indi/adhd200/}.
\newblock [Online; accessed 03-2013].

\bibitem[ADNI, 2013]{ADNI-url}
ADNI (2013).
\newblock Alzheimer's disease neuroimaging initiative.
\newblock \url{http://adni.loni.ucla.edu}.
\newblock [Online; accessed 03-2013].

\bibitem[Allen et~al., 2011]{Allen2011MatDecomp}
Allen, G., Grosenick, L., and Taylor, J. (2011).
\newblock A generalized least squares matrix decomposition.
\newblock {\em Rice University Technical Report No. TR2011-03},
  arXiv:1102:3074.

\bibitem[Aston and Kirch, 2012]{AstonKirch2012}
Aston, J.~A. and Kirch, C. (2012).
\newblock Estimation of the distribution of change-points with application to
  fmri data.
\newblock {\em Annals of Applied Statistics}, 6:1906--1948.

\bibitem[Blankertz et~al., 2001]{Blankertz2001}
Blankertz, B., Curio, G., and M{\"u}ller, K.-R. (2001).
\newblock Classifying single trial {EEG}: Towards brain computer interfacing.
\newblock In {\em NIPS}, pages 157--164.

\bibitem[Caffo et~al., 2010]{Caffo2010}
Caffo, B., Crainiceanu, C., Verduzco, G., Joel, S., S.H., M., Bassett, S., and
  Pekar, J. (2010).
\newblock Two-stage decompositions for the analysis of functional connectivity
  for {fMRI} with application to {A}lzheimer's disease risk.
\newblock {\em Neuroimage}, 51(3):1140--1149.

\bibitem[Chen et~al., 2001]{ChenDonohoSaunders01BasisPursuit}
Chen, S.~S., Donoho, D.~L., and Saunders, M.~A. (2001).
\newblock Atomic decomposition by basis pursuit.
\newblock {\em SIAM Rev.}, 43(1):129--159.

\bibitem[Crainiceanu et~al., 2011]{CrainiceanuCaffo11ImageDecomp}
Crainiceanu, C.~M., Caffo, B.~S., Luo, S., Zipunnikov, V.~M., and Punjabi,
  N.~M. (2011).
\newblock Population value decomposition, a framework for the analysis of image
  populations.
\newblock {\em J. Amer. Statist. Assoc.}, 106(495):775--790.

\bibitem[de~Leeuw, 1994]{deLeeuw94BR}
de~Leeuw, J. (1994).
\newblock Block-relaxation algorithms in statistics.
\newblock In {\em Information Systems and Data Analysis}, pages 308--325.
  Springer, Berlin.

\bibitem[Fan and Li, 2001]{FanLi01SCAD}
Fan, J. and Li, R. (2001).
\newblock Variable selection via nonconcave penalized likelihood and its oracle
  properties.
\newblock {\em J. Amer. Statist. Assoc.}, 96(456):1348--1360.

\bibitem[Frank and Friedman, 1993]{FrankFriedman93Bridge}
Frank, I.~E. and Friedman, J.~H. (1993).
\newblock A statistical view of some chemometrics regression tools.
\newblock {\em Technometrics}, 35(2):109--135.

\bibitem[Haxby et~al., 2001]{Haxby2001}
Haxby, J.~V., Gobbini, M.~I., Furey, M.~L., Ishai, A., Schouten, J.~L., and
  Pietrini, P. (2001).
\newblock Distributed and overlapping representations of faces and objects in
  ventral temporal cortex.
\newblock {\em Science}, 293(5539):2425--2430.

\bibitem[Hoff, 2011]{Hoff2011}
Hoff, P. (2011).
\newblock Hierarchical multilinear models for multiway data.
\newblock {\em Computational Statistics and Data Analysis}, 55:530--543.

\bibitem[Kolda and Bader, 2009]{KoldaBader09Tensor}
Kolda, T.~G. and Bader, B.~W. (2009).
\newblock Tensor decompositions and applications.
\newblock {\em SIAM Rev.}, 51(3):455--500.

\bibitem[Kontos et~al., 2003]{Kontos2003}
Kontos, D., Megalooikonomou, V., Kontos, D., Faloutsos, C., Megalooikonomou,
  V., Ghubade, N., and Faloutsos, C. (2003).
\newblock Detecting discriminative functional {MRI} activation patterns using
  space filling curves.
\newblock In {\em in Proc. of the 25th Annual International Conference of the
  IEEE Engineering in Medicine and Biology Society (EMBC}, pages 963--967.
  Springer-Verlag.

\bibitem[LaConte et~al., 2005]{Laconte2005}
LaConte, S., Strother, S., Cherkassky, V., Anderson, J., and Hu, X. (2005).
\newblock {{S}upport vector machines for temporal classification of block
  design f{M}{R}{I} data}.
\newblock {\em Neuroimage}, 26:317--329.

\bibitem[Lange, 2010]{Lange10NumAnalBook}
Lange, K. (2010).
\newblock {\em Numerical Analysis for Statisticians}.
\newblock Statistics and Computing. Springer, New York, second edition.

\bibitem[Lehmann and Romano, 2005]{lehmannRomano05TSH}
Lehmann, E.~L. and Romano, J.~P. (2005).
\newblock {\em Testing Statistical Hypotheses}.
\newblock Springer Texts in Statistics. Springer, New York, third edition.

\bibitem[McCullagh and Nelder, 1983]{McCullaghNelder83GLMBook}
McCullagh, P. and Nelder, J.~A. (1983).
\newblock {\em Generalized Linear Models}.
\newblock Monographs on Statistics and Applied Probability. Chapman \& Hall,
  London.

\bibitem[Mckeown et~al., 1998]{McKeown1998}
Mckeown, M.~J., Makeig, S., Brown, G.~G., Jung, T.-P., Kindermann, S.~S.,
  Kindermann, R.~S., Bell, A.~J., and Sejnowski, T.~J. (1998).
\newblock Analysis of {fMRI} data by blind separation into independent spatial
  components.
\newblock {\em Human Brain Mapping}, 6:160--188.

\bibitem[Mitchell et~al., 2004]{Mitchell2004}
Mitchell, T.~M., Hutchinson, R., Niculescu, R.~S., Pereira, F., Wang, X., Just,
  M., and Newman, S. (2004).
\newblock Learning to decode cognitive states from brain images.
\newblock {\em Machine Learning}, 57:145--175.

\bibitem[Qiu, 2005]{Qiu2005book}
Qiu, P. (2005).
\newblock {\em Image Processing and Jump Regression Analysis}.
\newblock Wiley series in probability and statistics. John Wiley.

\bibitem[Qiu, 2007]{Qiu07jumpsurface}
Qiu, P. (2007).
\newblock Jump surface estimation, edge detection, and image restoration.
\newblock {\em Journal of the American Statistical Association}, 102:745--756.

\bibitem[Ramsay and Silverman, 2005]{RamsaySilverman05FDABook}
Ramsay, J.~O. and Silverman, B.~W. (2005).
\newblock {\em Functional Data Analysis}.
\newblock Springer-Verlag, New York.

\bibitem[Reiss and Ogden, 2010]{ReissOgden10FunctionalGLM}
Reiss, P. and Ogden, R. (2010).
\newblock Functional generalized linear models with images as predictors.
\newblock {\em Biometrics}, 66:61--69.

\bibitem[Shinkareva et~al., 2006]{Ombao2006}
Shinkareva, S.~V., Ombao, H.~C., Sutton, B.~P., Mohanty, A., and Miller, G.~A.
  (2006).
\newblock Classification of functional brain images with a spatio-temporal
  dissimilarity map.
\newblock {\em NeuroImage}, 33(1):63--71.

\bibitem[Tibshirani, 1996]{Tibshirani96Lasso}
Tibshirani, R. (1996).
\newblock Regression shrinkage and selection via the lasso.
\newblock {\em J. Roy. Statist. Soc. Ser. B}, 58(1):267--288.

\bibitem[van~der Vaart, 1998]{vanderVaart98Asymp}
van~der Vaart, A.~W. (1998).
\newblock {\em Asymptotic Statistics}, volume~3 of {\em Cambridge Series in
  Statistical and Probabilistic Mathematics}.
\newblock Cambridge University Press, Cambridge.

\bibitem[Zhang, 2010]{Zhang10MCP}
Zhang, C.-H. (2010).
\newblock Nearly unbiased variable selection under minimax concave penalty.
\newblock {\em Ann. Statist.}, 38(2):894--942.

\bibitem[Zhou et~al., 2011]{ZhouArmagan11SparsePath}
Zhou, H., Armagan, A., and Dunson, D. (2011).
\newblock Path following and empirical {B}ayes model selection for sparse
  regressions.
\newblock {\em arXiv:1201.3528}.

\bibitem[Zhou et~al., 2013]{ZhouLiZhu12CPTensor}
Zhou, H., Li, L., and Zhu, H. (2013).
\newblock Tensor regression with applications in neuroimaging data analysis.
\newblock {\em Journal of the American Statistical Association}, In
  press(arXiv:1203.3209).

\bibitem[Zou and Hastie, 2005]{ZouHastie05Enet}
Zou, H. and Hastie, T. (2005).
\newblock Regularization and variable selection via the elastic net.
\newblock {\em J. R. Stat. Soc. Ser. B Stat. Methodol.}, 67(2):301--320.

\end{thebibliography}
\bibliographystyle{apalike}

\end{document}